%% file: main.tex
\documentclass[pmlr,twocolumn,10pt]{jmlr} %

\usepackage{booktabs}
\usepackage{siunitx}

\usepackage[switch]{lineno}

\usepackage{algorithmic}
\usepackage{graphicx}
\usepackage{textcomp}
\usepackage{booktabs}
\usepackage{siunitx}
\usepackage{bbm}
\usepackage{svg}
\usepackage{adjustbox}
\usepackage{cleveref}
\usepackage{hyperref}
\usepackage{caption}
\usepackage{multirow}
\usepackage{pgfplots}
\usetikzlibrary{patterns}
\pgfplotsset{compat=1.17}

\usepackage{xcolor}
\definecolor{lightblue}{HTML}{75b1eb}
\definecolor{lightyellow}{HTML}{e8b963}
\definecolor{lightgreen}{HTML}{81bf9d}
\definecolor{salmon}{HTML}{ed9d8e}

\DeclareOldFontCommand{\bf}{\normalfont\bfseries}{\mathbf}
\DeclareOldFontCommand{\it}{\normalfont\itshape}{\mathit}

\theorembodyfont{\upshape}
\theoremheaderfont{\scshape}
\theorempostheader{:}
\theoremsep{\newline}

\jmlrvolume{259}
\jmlryear{2024}
\jmlrsubmitted{LEAVE UNSET}
\jmlrpublished{LEAVE UNSET}
\jmlrworkshop{Machine Learning for Health (ML4H) 2024} %

 \title[HIST-AID: Leveraging Historical Patient Reports for Enhanced Multi-Modal Automated Diagnosis]{HIST-AID: Leveraging Historical Patient Reports for \\ Enhanced Multi-Modal Automatic Diagnosis}

\author{%
  \Name{Haoxu Huang}$^{1,3}$ \Email{hh2740@nyu.edu}\\
  \Name{Cem M. Deniz}$^{2}$ \Email{cem.deniz@nyulangone.org}\\
  \Name{Kyunghyun Cho}$^{1,3,4, 5}$ \Email{kyunghyun.cho@nyu.edu}\\
  \Name{Sumit Chopra}$^{1,2,3}$ \Email{Sumit.Chopra@nyulangone.org}\\
  \Name{Divyam Madaan}$^{1}$ \Email{divyam.madaan@nyu.edu}\\
  \addr $^1$Courant Institute of Mathematical Sciences, New York University\\
  \addr $^2$Department of Radiology, New York University Grossman School of Medicine\\
  \addr $^3$Center of Data Science, New York University\\
  \addr $^4$Prescient Design, Genentech \\ 
  \addr $^5$CIFAR LMB
}

\begin{document}

\maketitle

\begin{abstract}
Chest X-ray imaging is a widely accessible and non-invasive diagnostic tool for detecting thoracic abnormalities. While numerous AI models assist radiologists in interpreting these images, most overlook patients' historical data. To bridge this gap, we introduce \emph{Temporal MIMIC} dataset, which integrates five years of patient history, including radiographic scans and reports from MIMIC-CXR and MIMIC-IV, encompassing $12,221$ patients and thirteen pathologies.
Building on this, we present \emph{HIST-AID}, a framework that enhances automatic diagnostic accuracy using historical reports. HIST-AID emulates the radiologist's comprehensive approach, leveraging historical data to improve diagnostic accuracy. Our experiments demonstrate significant improvements, with AUROC increasing by $6.56\%$ and AUPRC by $9.51\%$ compared to models that rely solely on radiographic scans. These gains were consistently observed across diverse demographic groups, including variations in gender, age, and racial categories. We show that while recent data boost performance, older data may reduce accuracy due to changes in patient conditions. Our work paves the potential of incorporating historical data for more reliable automatic diagnosis, providing critical support for clinical decision-making. The code for generating the data and model training is available at \url{https://github.com/NoTody/HIST-AID}.
\end{abstract}

\section{Introduction}
\label{sec:intro}
\begin{figure*}[t!]
\centering
  \includegraphics[width=\linewidth]{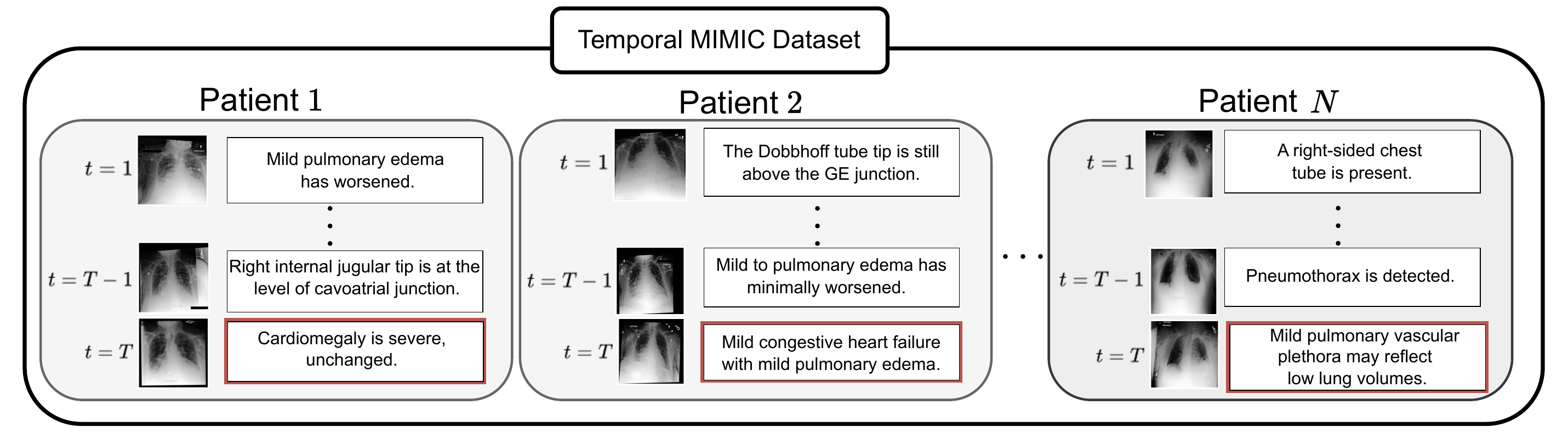}
   \caption{{\bf Temporal MIMIC Dataset:} The dataset consists of radiographic scans and corresponding radiology reports collected over a span of five years, providing a comprehensive view of the progression of patient conditions over time. The final report, highlighted in red, is used to obtain the ground-truth labels for the patient's current condition. \label{fig:dataset}}
    \includegraphics[width=0.8\linewidth]{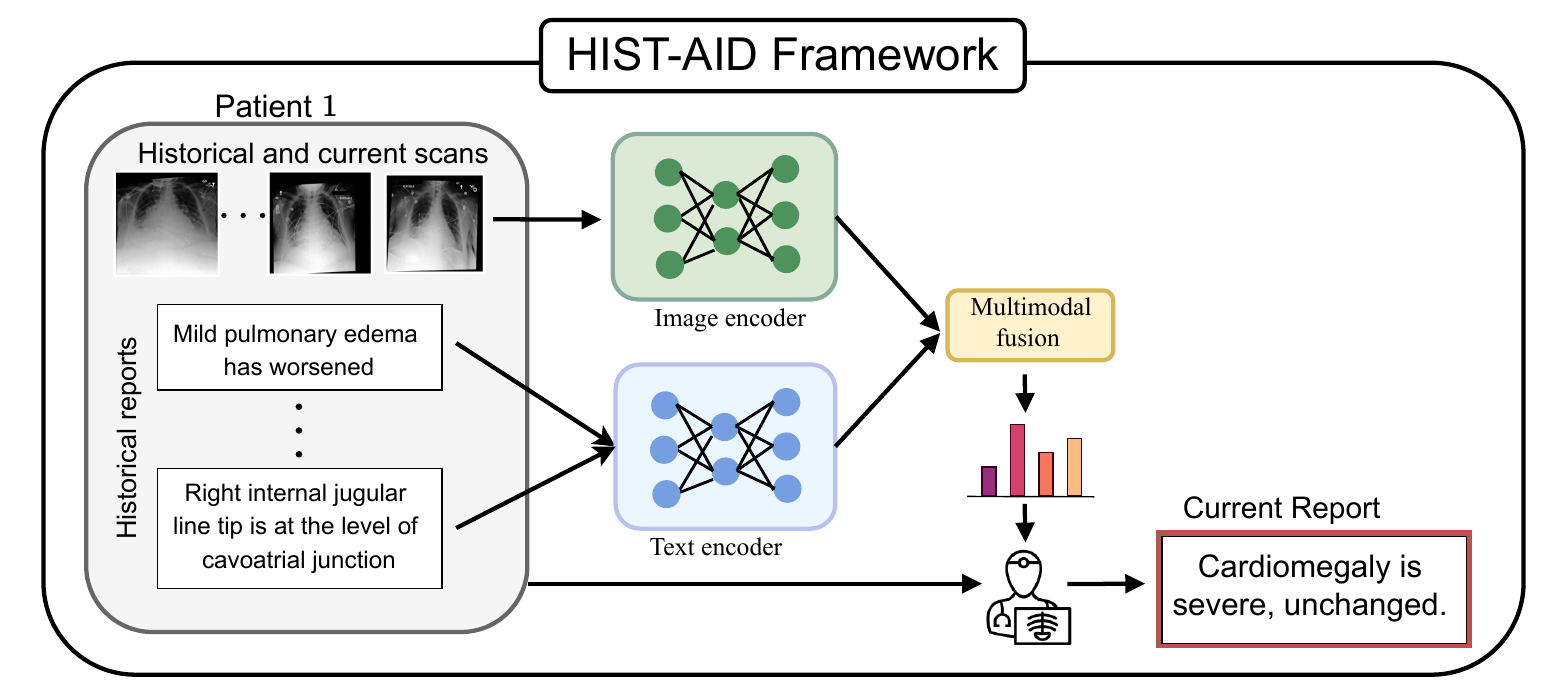}
\caption{{\bf Leveraging historical patient scans and reports for automatic diagnosis (HIST-AID) Framework:} We retrieve both current and historical image scans along with the radiology reports from the past. These inputs are first processed through image and text encoders. Subsequently, the resulting modality-specific representations are combined with time-series information (time offset from the current time stamp) to generate the final predictions with multi-modal fusion. This prediction, combined with the current scan and the patient's historical scans and reports, assists the physician in the final diagnosis.\label{fig:combined_figure}}
\end{figure*}

Chest X-ray (CXR) is widely used for diagnosing thoracic abnormalities due to its affordability, accessibility, and non-invasive nature~\citep{chestxray14, irvin2019chexpert, Johnson2019-mu, Johnson2023-wh, huang2023radiology}. AI-driven clinical decision support systems have shown potential to match or exceed human diagnostic accuracy~\citep{rajpurkar2018deep, killock2020ai, gaube2023non}. However, most deep learning models only focus on the latest scan, neglecting patients' historical data~\citep{chestxray14, chexnet, irvin2019chexpert, KHAN2020105581}. This oversight is a critical limitation, as radiologists  incorporate a patient’s medical history and track changes over time to provide a more accurate diagnosis.

To address this, we introduce the \emph{Temporal MIMIC} dataset (see \Cref{fig:dataset}), which provides a comprehensive longitudinal view of patient data by combining five years of radiology images from MIMIC-CXR~\citep{Johnson2019-mu} with corresponding clinical reports from MIMIC-IV~\citep{Johnson2023-wh}. This dataset includes $12,221$ patients, each with an average of eleven reports and thirteen scans, providing a rich temporal multi-modal dataset that facilitates the development of multi-modal models capable of detecting subtle changes in a patient's condition over time.

To fully leverage our proposed dataset, we propose \emph{HIST-AID}, an end-to-end framework that leverages historical chest X-rays and reports for abnormality detection (see \Cref{fig:combined_figure}). In clinical deployment, when a patient undergoes a Chest X-Ray, our framework retrieves and integrates the historical clinical data from the healthcare system database. HIST-AID uses transformer-based image and text time-series encoders~\citep{tst}, effectively capturing temporal information from these past scans and reports. The temporal information from past scans and reports are then combined through multi-modal fusion~\citep{pmlr-v139-kim21k} to make a preliminary diagnosis. The radiologists review and utilize this diagnosis to write the final report. By modeling these historical trends, HIST-AID can identify evolving patterns in a patient’s condition that may not be apparent from a single scan. 

Our evaluation shows that integrating past reports improves model performance across thirteen pathologies, with average AUROC and AUPRC increases of $6.56\%$ and $9.51\%$ compared to current scan only methods. This improvement is consistent across subgroups defined by gender, age, and race, ensuring a more equitable diagnostic approach. Incorporating past scans with reports did not yield additional gains, likely due to overlapping information between the two modalities for abnormality detection. Additionally, we found that reports from distant timestamps led to a decline in performance, highlighting that not all historical data is equally useful.

\section{Related work}\label{sec:related_works}
In this section, we provide an overview of the datasets for CXRs and radiology reports, along with the machine learning models developed using these datasets.

\paragraph{Datasets.}
Chest radiography is essential for early detection and diagnosis of critical health conditions, with recent deep learning advancements enhancing its effectiveness through large-scale datasets. Prominent datasets like ChestX-ray14~\citep{chestxray14} and CheXpert~\citep{irvin2019chexpert} provide extensive radiographic images with automated annotations, while the MIMIC dataset~\citep{Johnson2019-mu, Johnson2023-wh} offers a large number of CXRs along with Electronic Health Record (EHR), including timestamps, patient identifiers, and hospital admission data.  

Our work is the first to introduce a longitudinal dataset that combines historical CXRs from MIMIC-CXR~\citep{Johnson2019-mu} and corresponding clinical reports from MIMIC-IV~\citep{Johnson2023-wh}, allowing models to leverage historical medical data to improve diagnostic accuracy. 

\paragraph{Models.}
Numerous deep learning approaches have advanced pathology diagnosis prediction. Early works like CheXNet~\citep{chexnet} demonstrate performance comparable to experienced radiologists in Chest X-ray pathology prediction. Recent studies in multi-modal pre-trainin and leveraging images, text, and tabular data further improves predictive models. ConVIRT~\citep{zhang2022contrastive} enhances visual representations with image-text pairs, while BiomedCLIP~\citep{biomedclip} scales CLIP-style pre-training with biomedical data from PubMed. BioViL-T~\citep{bannur2023learning} explores image-text contrastive pre-training using historical and current images.

Recently, time-series modeling has also gained attention in the community.~\citet{Barbieri2020} evaluates RNNs on ICU readmission risk using electronic medical records, and \citet{Kaushik2020-ue} shows the effectiveness of ensembles and LSTMs in predicting healthcare costs. HAIM~\citep{Soenksen2022} integrates CXRs and reports along with time-series information but lacks end-to-end training and proper timestamp segregation, resulting in biases and data leakage. Our work incorporates both historical radiology notes and images by addressing issues like improper data partitioning and refining the overall problem setup to enable accurate, end-to-end diagnosis using patient history.

\section{Method}
In this section, we outline the problem setup for abnormality prediction using historical data, followed by the process of generating the Temporal MIMIC dataset. We then introduce our HIST-AID framework, which leverages this dataset to enhance diagnostic accuracy.

\subsection{Problem Setup}  
We consider a dataset of $N$ patients, represented as $\{({X}^{\text{image}}_i, {X}^{\text{text}}_i, y_i)\}_{i=1}^N$, where each patient $i$ is associated with a temporal sequence of imaging scans and corresponding clinical reports over time. The temporal sequence is indexed by ${T} = (1, \ldots, t_n)$, where $t_n$ denotes the latest timestamp for the imaging data.

The imaging data for each patient $i$ is expressed as ${X}^{\text{image}}_i = \{x^\text{image}_{i,t}\}_{t=1}^{t_n}$, where each $x^\text{image}_{i,t}$ represents an image from timestamp $t$. The corresponding sequence of textual reports is denoted as ${X}^{\text{text}}_i = \{x^\text{text}_{i,t}\}_{t=1}^{t_{i'}}$, with $t_{i'}$ typically being one timestamp prior to $t_n$. 

The objective is to predict the pathology label $y_i$ for the scan at the current timestamp $t_n$, leveraging all available imaging data and associated historical reports from previous timestamps.

\subsection{Temporal-MIMIC dataset generation}\hypertarget{sec:dataset}{}

The objective of Temporal MIMIC dataset is to enable the use the historical images and patient reports to evaluate their utility for automatic diagnosis. To accomplish this, we integrate Chest-X rays from MIMIC-CXR~\citep{Johnson2019-mu} and radiology reports from MIMIC-IV~\citep{Johnson2023-wh}, linked through patient subject identifiers.  

 Temporal MIMIC contains $12,221$ patients from $69,077$ radiographic studies, with an average of eleven reports and thirteen images per patient, paired with free-text radiology reports collected between 2011 and 2016 at Beth Israel Deaconess Medical Center in Boston, MA. Each data point is timestamped relative to the patient’s most recent radiology image and includes corresponding ground truth pathology labels. The dataset spans thirteen distinct multi-label pathologies, including atelectasis, cardiomegaly, edema, lung opacity, pleural effusion, pneumonia, and pneumothorax, as well as fewer instances of fractures, lung lesions, and other pleural disorders.

The construction of {Temporal-MIMIC} dataset consists of the the following steps. 
\begin{enumerate}
    \item {\bf Data merging.} We link the images from MIMIC-CXR~\citep{Johnson2019-mu} with corresponding reports drawn from the time-series patient records of MIMIC-IV~\citep{Johnson2023-wh}. This connection is facilitated by the common identifiers present in both datasets, with labels derived directly from the radiology reports.
    \item {\bf Remove current time-stamp.} We exclude the current timestamp data from the reports to simulate a real-world scenario where a diagnosis must be made without immediate access to the latest diagnostic information.
    \item {\bf Augmentation with additional patient samples.} We treat each valid timestamp during a patient's admission period as a separate sample, significantly increasing the number of samples per patient. Each timestamp is linked to the corresponding labels in MIMIC. For each timestamp, we create a new datapoint by combining the label from that timestamp with all previous timestamps. For instance, if a patient has five timestamps, we take the label from the fifth timestamp and combine it with the prior four timestamps to form one datapoint. Similarly, we take the label from the fourth timestamp and combine it with the preceding three timestamps as another datapoint, and so on.
    \item {\bf Removing duplicates.} Post-merging, any duplicate records identified within the historical patient data are removed to ensure dataset integrity and prevent redundancy in the training process. Entries with empty impression section are also removed in this step.
    \item {\bf Dataset splitting.} Finally, the dataset is divided into $80\%$  (n = $55,471$) for training, $10\%$ (n = $6,776$) for validation, and $10\%$ (n = $6,830$) for testing. This split is performed using unique subject IDs from the MIMIC dataset to ensure that all data points related to a single patient are contained within one subset, thereby avoiding potential data leakage across the different phases of model evaluation. 
\end{enumerate}

\Cref{fig:dataset} shows various samples from our dataset. %
\Cref{fig:main-dataset-figure} shows the construction of Temporal MIMIC dataset. We provide the details on demographic distribution of the study population, label distribution and co-occurrences in the supplementary material.

\subsection{HIST-AID framework}\hypertarget{sec:methods}{}
HIST-AID, shown in \Cref{fig:combined_figure}, leverages historical images and reports in a temporal model, mimicking the radiologist's workflow and improving diagnostic performance over using only the current scan.

We use distinct modality-specific encoders: $f^{\text{image}}_{\theta}$ and $f^{\text{text}}_{\phi}$ for CXRs and corresponding reports from different time-stamps. These time-series representations are then processed by a separate time series encoder for each modality. The output of these encoders is followed by aggregation using multi-modal fusion encoder $h_{\tau}$. 
The predicted pathology output $\widehat{y}$ leveraging both the historical imaging scans and associated reports as follow:
\begin{equation}\label{eq:main_eq}
{\widehat{y}}=h_{\tau}\left(\bigoplus_{i} f^{\text{image}}_{\theta}\left({X}^{\text{image}}_{i}\right), \bigoplus_{i} f^{\text{text}}_{\phi}\left({X}^{\text{text}}_{i}\right)\right)
\end{equation}
The aggregation operation \( \bigoplus \) means the separate representations from different timestamps are concatenated to form a sequence of representations, where the sequence is padded by zero vectors if the sequence length is shorter than pre-defined maximum sequence length. We discuss the components in greater detail below.

\subsubsection{Pre-trained encoders}
In numerous studies on medical data~\citep{sowrirajan2021moco, radimagnet, wang-etal-2022-medclip, eslami2023pubmedclip, biomedclip}, models pre-trained on relevant datasets consistently outperformed baseline models that were not pre-trained. In our work, CXRs and reports are encoded by pre-trained modality-specific encoders: a vision transformer (ViT)~\citep{dosovitskiy2020vit} for images and a BERT encoder ~\citep{devlin-etal-2019-bert} for text, both from BiomedCLIP~\citep{biomedclip}.

These encoders are denoted as $f^\text{image}_\theta$ and $f^\text{text}_\phi$ and are used to process historical radiology images ${X}^{\text{image}}_{i}$ and reports ${X}^{\text{text}}_{i}$ respectively (see \Cref{eq:main_eq}). We use the embedding of the $[CLS]$ token and append zero to the last dimension to accommodate time-series data that is shorter than the maximum length.

\subsubsection{Time-series and multi-modal encoder}

To effectively capture longitudinal information across different timestamps, our approach uses a transformer encoder for both time-series and multi-modal inputs, rather than averaging representations across timestamps~\citep{Soenksen2022}. The time-series and multi-modal inputs are encoded into $\{{X}^{\text{image}}_{i},{X}^{\text{text}}_{i}\} \in \mathbb{R}^{B \times K \times D}$, where $B$ is the batch size, $K$ is the maximum time-series length, and $D$ is the output dimension of the encoder using the $[CLS]$ token embedding for each image or text encoder. Learnable tokens, [IMG] for images and [TEXT] for text, are added to each representation at each timestamp. If the time-series length is less than $K$, zero padding is applied.

For simplicity, assume the maximum image timestamp is $T$ and the maximum text timestamp is $T-1$. We define $f_\phi\left(x^{\text{modality}}_{i, t_j}\right) \in \mathbb{R}^{1 \times d}$ as the output for each modality (image or text), and aggregate the outputs across timestamps as $\bigoplus f^{\text{modality}}_\phi\left({X}_i\right) \in \mathbb{R}^{K \times d}$. The concatenated outputs, along with the $[CLS]$ token, form the transformer input with a shape of $\mathbb{R}^{2T}$.

\input{results_figure}
\input{demographics_figure}

We employ Rotary Positional Encoding (RoPE)~\citep{rope} to encode time-series information. We discuss the effectiveness of RoPE over other positional encoding methods in the supplementary material. A min-max normalized time offset serves as the positional indicator within the time-series. Given time offsets $\mathbf{t} = \{t_1, ..., t_n\}$, the normalized offset at timestamp $t_i$ is:
\begin{equation}
t_i^{\text{norm}} = \frac{t_i - \min(t_1, ..., t_n)}{\max(t_1, ..., t_n) - \min(t_1, ..., t_n)}
\end{equation}
$[CLS]$ token is added to the multi-modal representations as $\{[CLS], {X}^{\text{image}}_{i}, {X}^{\text{text}}_{i}\}\in \mathbb{R}^{B \times K \times (D+1)}$, capturing holistic information across modalities. This aggregated representation is encoded using a time-series transformer~\citep{tst} for early fusion, and the encoded output is fed into a linear classifier for pathology classification.

\begin{figure*}[ht!]
    \centering
    \begin{minipage}{0.45\textwidth}
        \centering
        \includegraphics[width=\linewidth]{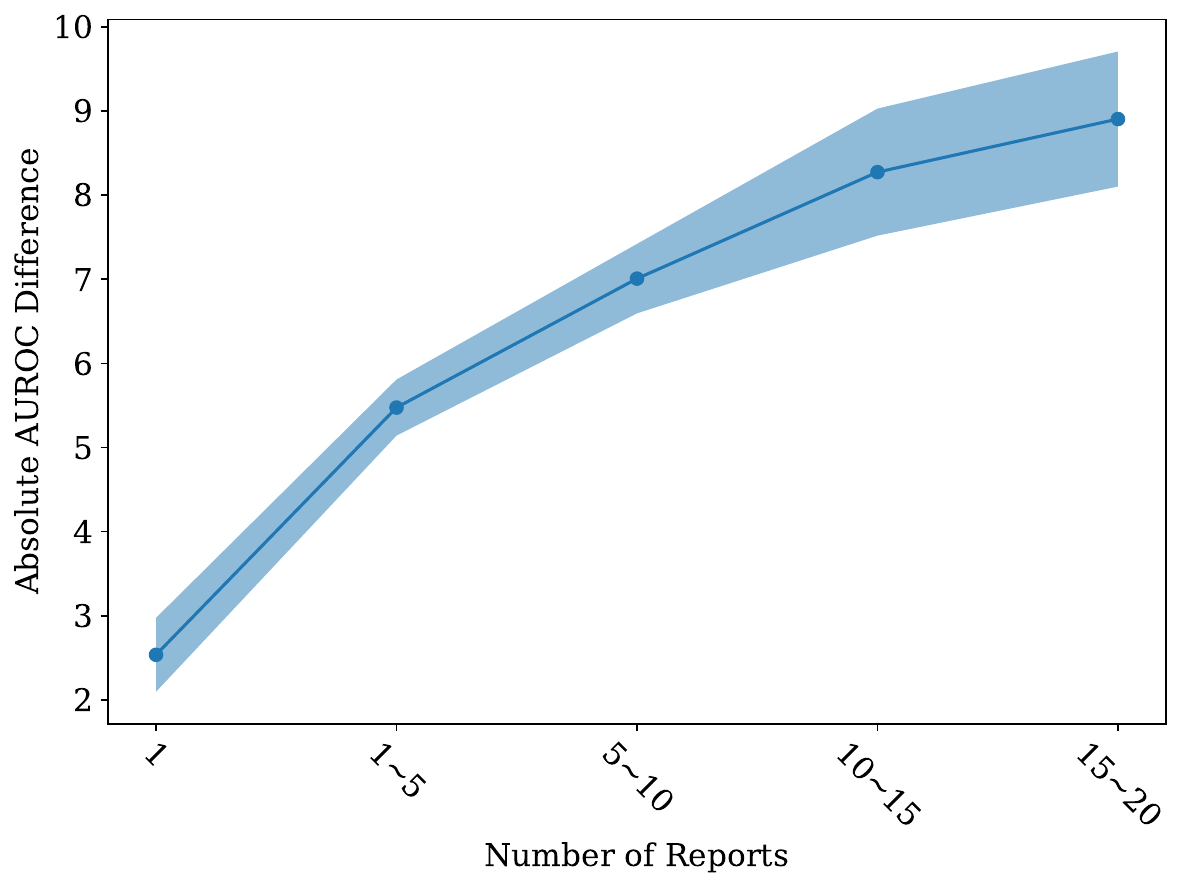}
        \caption{{\bf Impact of increasing the number of reports on AUROC performance:} The performance of the temporal multi-modal model enhances as the number of reports increases, surpassing the model that relies solely on current timestamp images.}
        \label{fig:num_reports}
    \end{minipage}
    \hfill
    \begin{minipage}{0.45\textwidth}
        \centering
        \includegraphics[width=\linewidth]{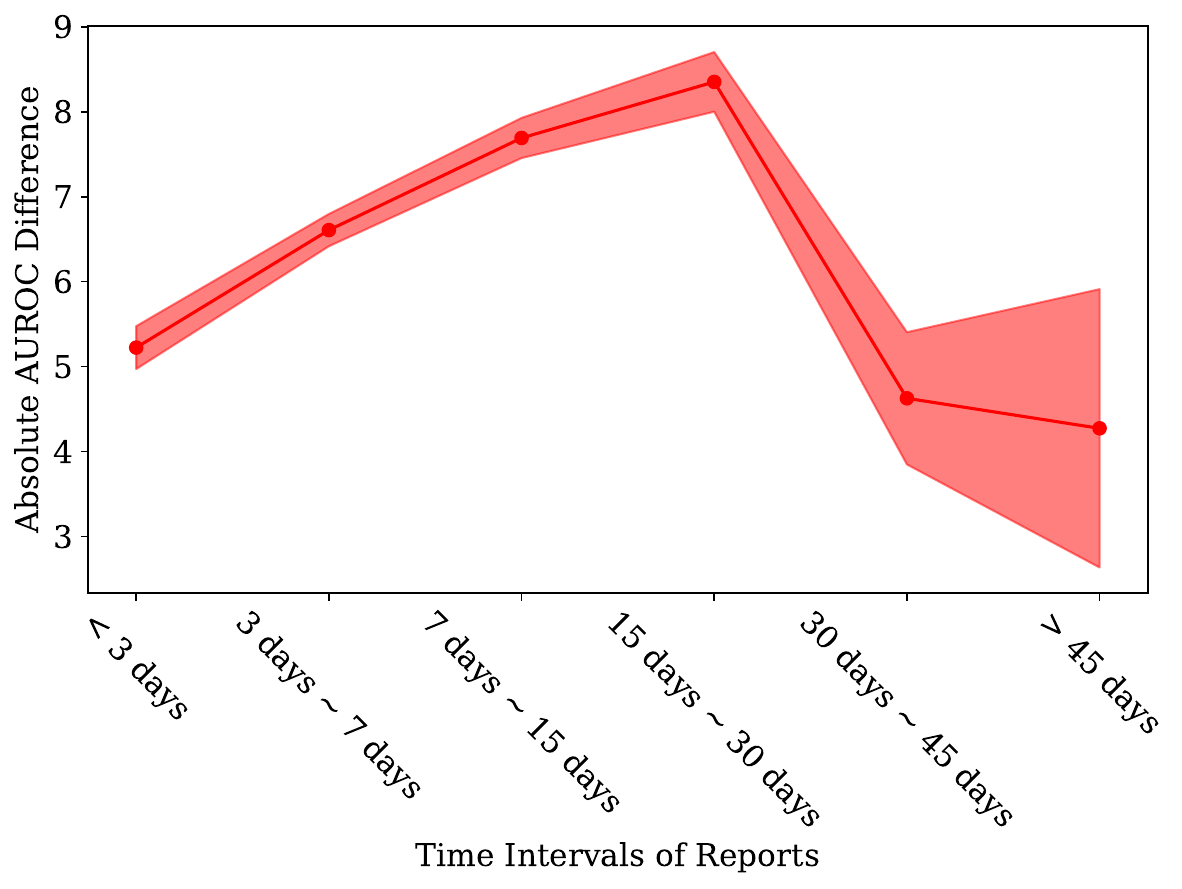}
        \caption{{\bf Impact of report timing relative to final diagnosis on AUROC performance:} The performance of the temporal multi-modal model improves when utilizing reports from up to the last 30 days, but it declines with reports from more distant periods, cautioning against the use of older data.}
        \label{fig:time_intervals_reports}
    \end{minipage}
\end{figure*}

\section{Results}

We compare the performance of HIST-AID on the Temporal MIMIC dataset against a uni-modal model that uses only the current scan~\citep{chexnet, chestxray14, irvin2019chexpert}. In addition, we analyze the performance across different demographic subgroups, the impact of increasing the number of historical reports and the influence of reports from different time periods. All models are trained with multi-label classification with mean and standard deviation over five
runs. Further details about the dataset and hyper-parameters for the models are deferred to the supplementary material. 

\subsection{Temporal multi-modal learning improves pathology prediction}
The incorporation of historical radiology reports with our temporal multi-modal model during fine-turning shows a significant improvement in AUROC performance when compared to models trained with scans from the current time-stamp for all the examined pathologies in \Cref{fig:main_eval}. We observe over $5\%$ improvement in AUROC in specific pathologies such as Consolidation ($5.05\%, p<0.0001$), Pleural Other ($5.55\%, p<0.0001$), Pneumonia ($6.20\%, p<0.0001$). On average, HIST-AID shows an enhancement of $6.56\%~(p<0.0001)$ and $3.41\%~(p<0.0001)$ in AUROC for all pathologies when compared to models relying solely on chest radiographs and historical reports respectively. Additionally, we observe a $9.51\%~(p<0.0001)$ and $2.63\%~(p<0.0001)$ improvement in average AUPRC (see supplementary material) for all the pathologies.
These results confirm the advantage of using CXRs with the historical reports for detecting thoraic abnormalities.

\subsection{Subcohort analysis}

We compare the performance of the model trained using only the current scan to HIST-AID across various demographic sub-cohorts, including gender, age, and race, as shown in \Cref{fig:demo_perf}. This analysis was inspired by previous research highlighting significant disparities in model fairness and effectiveness when applied to diverse groups~\citep{Seyyed-Kalantari2021, zhang2022improving, Yang2024, Vaidya2024}. Our results show a clear advantage for HIST-AID over the current scan only approach across all demographic subgroups.

HIST-AID consistently demonstrates accurate pathology diagnosis across both male and female patients, suggesting it helps reduce gender disparities. For age, the model achieved higher AUROC for younger and middle-aged adults, though performance slightly declined for individuals over 60, potentially due to the increased complexity of disease manifestations in older patients. While HIST-AID performed slightly less effectively for the black population compared to other racial groups, it provided a substantial improvement of $6-7\%$ AUROC compared to the current scan only model across all racial demographics. 

These findings highlight the potential of HIST-AID to enhance diagnostic accuracy across diverse demographic groups, helping mitigate biases present in models trained with only the current scan.

\subsection{Enhancing pathology prediction with additional radiology reports}
To assess the impact of the number of historical reports on automatic diagnosis, we measure the absolute AUROC improvement between models trained with both CXRs and historical reports and those trained with CXRs alone in \Cref{fig:num_reports}. We observe positive correlation between the number of reports and the AUROC improvement, indicating that incorporating more historical radiology reports improves diagnostic performance. Detailed breakdown across pathologies is in the supplementary material.

\Cref{fig:time_intervals_reports} shows the temporal relevance of the reports by examining the impact of time intervals between the reports and the final diagnosis on AUROC. This analysis is crucial, as the number of reports does not necessarily correlate with their time distribution—multiple reports may originate from a single time period. We observe consistent performance improvement when reports are within 30 days of the diagnosis, while older reports tend to reduce AUROC. This highlights the importance of recent information in the diagnostic process and suggests caution when including older reports,

\begin{table}
\resizebox{\linewidth}{!}{%
    \begin{tabular}{lccc}
    \toprule
    \textbf{Methods} & \multicolumn{2}{c}{\textsc{FT}}{\textsc{FLOPS}} \\
    \midrule
    \textsc{ConcatMLP} & 75.00$\pm$0.05 & 9.69G & \\
    \textsc{Block \citep{block}} & 76.26$\pm$0.21 & 10.02G & \\
    \textsc{MBT \citep{nagrani2021attention}} & 74.46$\pm$0.04 & 9.56G & \\
    \textsc{ViLT \citep{pmlr-v139-kim21k}} & \textbf{77.49}$\pm$0.14 & 9.38G & \\
    \textsc{METER \citep{METER}} & 75.88$\pm$0.11 & 12.50G & \\
    \bottomrule
    \end{tabular}
}
\captionof{table}{{\bf Ablation on Different Fusion Methods.} Mean AUROC across all pathologies is reported in this study. We observe that ViLT represents the best trade-offs between model performance and compute.}
\label{tab:ablation-fusion}
\end{table}

\subsection{Effect of fusion methods}
We show model performance on different multi-modal fusion methods in \Cref{tab:ablation-fusion} with their detailed descriptions in the supplementary material. Our analysis highlights that ViLT~\citep{pmlr-v139-kim21k} (early fusion) achieves an optimal balance between performance and computational efficiency within our framework. ViLT concatenates the representation tokens from image and historical reports and uses them as input to the time-series transformer encoder for information extraction. This demonstrates that early fusion is particularly effective for medical diagnosis tasks involving image and text modalities, where inter-modality dependencies~\citep{madaan2024jointly} are important. By integrating modality interactions from the outset, ViLT captures cross-modal relationships more effectively. Moreover, it avoids the need for separate time-series encoders for different modalities, making it more computationally efficient compared to alternative methods.

\input{ablations_figure}

\section{Limitations and future work}

\paragraph{Challenges in leveraging historical radiographic scans for diagnostic performance.}
While our objective was to enhance diagnostic accuracy by incorporating both historical scans and reports, emulating the workflow of radiologists, our findings reveal limitations in the effectiveness of historical radiographic scans. \Cref{fig:hist_abla} compares models relying on the latest images and reports with those utilizing historical information in both uni-modal and multi-modal settings, with mean values and $95\%$ confidence intervals across five independent runs.

In the uni-modal setting, incorporating historical CXRs and reports improves performance independently as expected. In the multi-modal setting, while historical reports boost diagnostic accuracy, adding historical images does not yield similar benefits. This limitation may stem from the fact that radiologists typically extract key pathology information from current images and document it in reports, making information from historical texts overlap with that of corresponding images.

Additionally, we hypothesize that optimization challenges in integrating high-dimensional historical scans into multi-modal models contribute to this issue. Addressing these challenges and developing more effective end-to-end multi-modal training techniques for incorporating historical scans is an important direction for future research.

\paragraph{Performance trade-offs in using different report sections.}

A limitation we identified is the varying contribution of different sections of radiology reports to model performance. As AI-based diagnostic tools evolve, understanding the impact of specific report sections becomes increasingly important, as these tools may influence how clinicians structure their reports. Radiology reports typically consist of five distinct sections, each serving a unique function:

\begin{itemize}
\item {\bf History.} Provides a brief overview of the patient’s medical background.
\item {\bf Indication.} Lists the reasons for conducting the radiological procedure.
\item {\bf Comparison.} References previous scans for comparison with the current one.
\item {\bf Findings.} Contains detailed observations made by the radiologist in each scanned area.
\item {\bf Impressions} Summarizes the key findings and their potential implications.
\end{itemize}

While all sections of the reports are important, many are frequently incomplete or missing. For instance, the history and comparison sections are often absent or lack sufficient detail. As a result, we focus on the findings and impression sections, as they contain the most critical diagnostic information and are more consistently present. Specifically, we analyze the impact of these sections on model performance to better understand their contribution. These sections are combined using the template: ``Impression: $\langle$\textit{impression text}$\rangle$ Finding: $\langle$\textit{finding text}$\rangle$''. As shown in \Cref{fig:sections_abla}, including both sections improves performance by $0.69\%$ ($p < 0.05$). However, this comes at a significant computational cost, as the findings section is two to three times longer than the impression section in terms of token length, as detailed in the supplementary material.
Reducing training time and developing efficient models using all report sections is a promising direction for future research.

\section{Conclusion}
In this paper, we introduced the Temporal-MIMIC dataset, designed to assess model performance by integrating radiology images and reports across a patient’s medical history. To leverage these historical images and reports for automated diagnosis, we proposed HIST-AID, a multi-modal framework that encodes modality-specific representations of both images and text, which are then combined through multi-modal fusion. 
Our results demonstrated that incorporating historical radiological reports alongside current scans significantly enhances the accuracy of automatic abnormality detection in chest X-rays, delivering consistent improvements across subgroups defined by gender, age, and race, thereby promoting a more equitable diagnostic approach. We showed that the impact of historical data varies across time, with the most recent reports -- upto 30 days from diagnosis -- being valuable, while older records tend to diminish predictive performance. This underscores the importance of carefully selecting relevant time periods when utilizing past medical information.
By leveraging historical patient records, HIST-AID will enable specialists to comprehensively model patient histories, facilitating more effective identification of high-risk patients. This approach will help us to transform care delivery, improve treatment outcomes, and enhance overall healthcare efficiency.

\acks{We thank the anonymous reviewers for their insightful comments and suggestions. This research was supported by Samsung Advanced Institute of Technology (Next Generation Deep Learning: from pattern recognition to AI), Hyundai NGV (Uncertainty in Neural Sequence Modeling in the Context of Dialogue Modeling), NSF Award 1922658, the Center for Advanced Imaging Innovation and Research (CAI2R), a National Center for Biomedical Imaging and Bioengineering operated by NYU Langone Health, National Institute of Biomedical Imaging and Bioengineering through award number P41EB017183. The computational requirements for this work were supported in part by NYU IT High Performance Computing resources, services, and staff expertise and NYU Langone High Performance Computing Core's resources and personnel. This content is solely the responsibility of the authors and does not represent the views of the funding agencies.
}

\bibliography{references}

\clearpage
\appendix
\renewcommand\thefigure{\thesection.\arabic{figure}}    
\renewcommand\thetable{\thesection.\arabic{table}}

\section{Model Implementation and Training Details}

\subsection{Statistical Analysis}
One-tailed Wilcoxon rank-sum test ($\alpha=0.05$) was used to compute all p-values reported in the paper, where the samples are drawn from models trained on 5 different seeds to evaluate the stability of the models and robustness of the hypothesis.

\subsection{Data Augmentations}
For radiology images, we apply Random Resized Crop with scale $0.6-1.0$ and Bicubic Interpolation, Color Jittering with brightness $0.4-0.6$, contrast $0.4-0.6$, no saturation and hue change with probability $0.5$, and Random Horizontal Flip with probability $0.5$. We do not apply any data augmentations on radiology reports.

\subsection{Hyperparameters}
\begin{figure*}[!t]
\centering
\begin{minipage}{0.5\linewidth}
\centering
\resizebox{\linewidth}{!}{
\begin{tabular}{lllll|l}
\hline
Hyper-Parameter  & \multicolumn{2}{l}{Value}  \\ 
\hline
Batch Size    & 16 &  \\
Learning Rate (Image Encoder)   & 1e-5 $\times$ (Batch Size / 64) &  \\
Learning Rate (Text Encoder)   & 1e-5 $\times$ (Batch Size / 64) &  \\
Learning Rate (Time Series Encoder)   & 1e-4 $\times$ (Batch Size / 64) &  \\
Epochs              &    15  &  \\
Weight Decay  &    1e-2  &  \\
Optimizer    &  AdamW  & \\
AdamW Betas        &    (0.9, 0.999)  &  \\
Scheduler     &    Cosine with Linear Warmup  & \\
Linear Warmup Steps  &   0.10 $\times$ Total Training Steps & \\
Minimum Learning Rate    &  1e-3 $\times$ Learning Rate & \\
Image Time-Series Max Sequence Length   &  1  & \\
Text Tokens Max Length    & 200 & \\
Text Time-Series Max Sequence Length   &  50  & \\
\hline
\end{tabular}
}
\captionof{table}{Hyper-parameters for Model Training}
\label{tab:hyperparam}
\end{minipage}%
\hfill
\begin{minipage}{0.48\textwidth}
  \includegraphics[width=\textwidth]{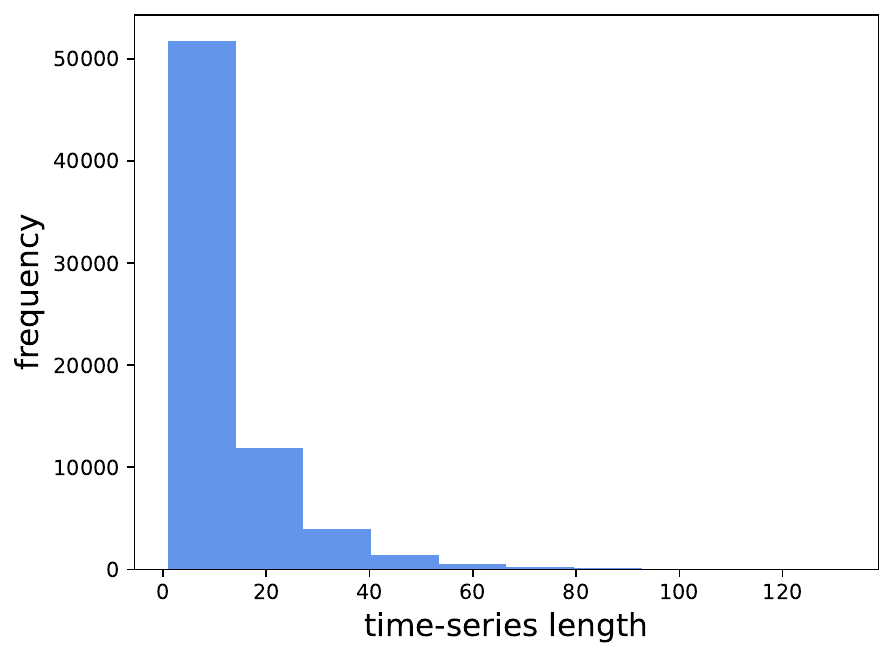}
  \caption{Distribution for Text Time-Series Length}
  \label{fig:text_time_series}
\end{minipage}
\end{figure*}

Table \ref{tab:hyperparam} provides hyperparameters for training our framework. We keep the same hyperparameters in uni-modal cases. We performed a small-scale hyperparameters search to ensure our result does not change too much on different hyperparameters settings. For experiments of using finding section or both impression and finding sections, we use 400 Text Tokens Max Length due to computational constraint.

We show the distribution of time-series length for all samples in Figure \ref{fig:text_time_series}, where we select max sequence padding with truncation to 50 under evaluation of computational cost and length coverage.

\subsection{Training Details}
We use 12 layers Vision Transformer (ViT-Base) with path size $16\times 16$ \citep{dosovitskiy2020vit} as Image Encoder and 12 layers BERT-Base \citep{devlin-etal-2019-bert} as Text Encoder in all our experiments. The pre-trained weights of both encoders are loaded from BiomedCLIP \citep{biomedclip}. We use the model checkpoint with best validation AUROC for testset performance evaluation. The validation performance is calculated after each epoch. We do not use mixed-precision training as we find the training to be unstable with mixed-precision. All experiments are performed on two NVIDIA A100 80GB GPUs with total training time range from approximately 15 to 20 hrs varied based on different settings.

\subsection{Dataset}
\begin{figure*}[!t]
    \centering
    \includegraphics[width=\linewidth]{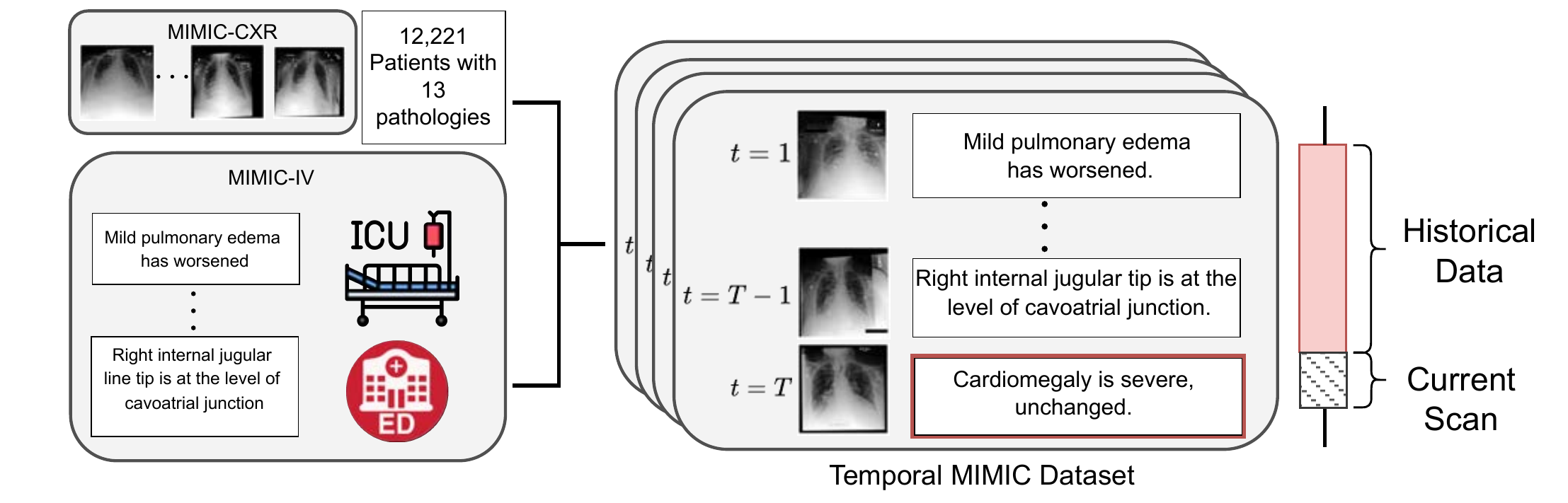}
    \caption{{\bf Overview of Temporal MIMIC creation:} We integrate  historical and current Chest X-rays from MIMIC-CXR with radiology reports from MIMIC-IV using unique patient identifiers.}
    \vspace{0.2in}
    \label{fig:main-dataset-figure}
\end{figure*}

\begin{figure*}[!t]
    \centering
    \begin{minipage}{0.45\linewidth}
        \begin{tabular}{llr}
            \toprule
            \multicolumn{2}{c}{Characteristics} & Count (Proportion \%) \\
            \midrule
            \multirow{4}{*}{Race}   & Asian & 2,269 (3.3\%) \\
                                    & Black & 8,147 (11.8\%) \\
                                    & White & 45,637  (66.1\%)\\
                                    & Other & 13,024 (18.8\%) \\
            \midrule
            \multirow{2}{*}{Sex}    & Female & 29,021 (42.0\%) \\
                                    & Male & 40,056 (58.0\%) \\
            \midrule
            \multirow{4}{*}{Age}    & 0-40 & 6,597 (9.6\%) \\
                                    & 40-60 & 20,461 (29.6\%) \\
                                    & 60-80 & 32,039 (46.4\%) \\
                                    & 80-100 & 9,980 (14.4\%) \\
            \bottomrule
        \end{tabular}
                \centering
        \captionof{table}{{\bf Demographic distribution of the study population.} The demographic breakdown reveals a predominantly white cohort, a slightly male-dominated gender ratio. Additionally, there is a significant segment of participants aged 40-80.
        \label{tab:demographics_tab}}
    \end{minipage}%
    \hfill
    \begin{minipage}{0.53\linewidth}
        \centering
        \includegraphics[width=\linewidth]{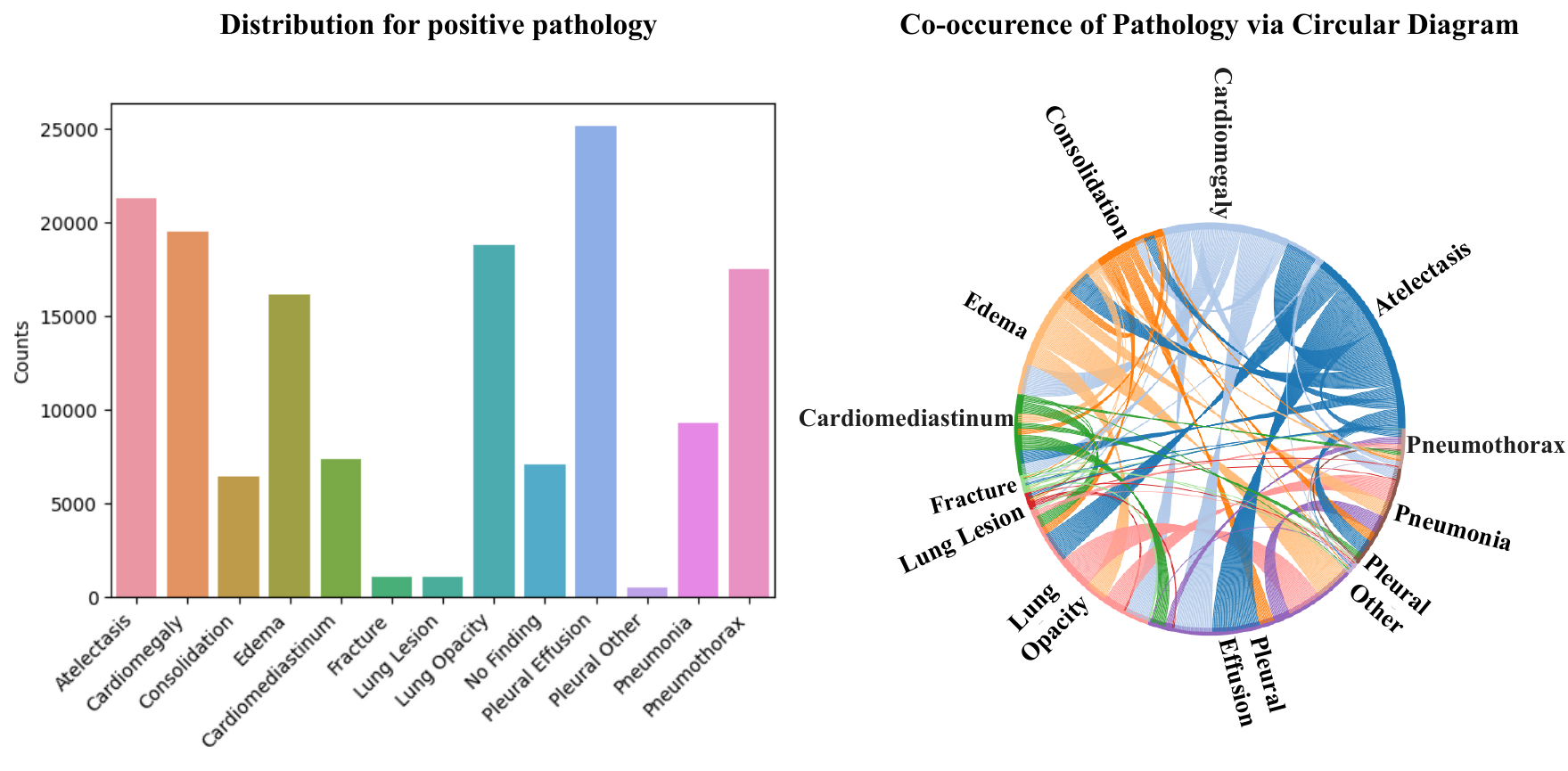}
        \vspace{-0.2in}
    \captionof{figure}{{\bf Distribution and co-occurrence of various pathologies.} The left bar graph representing the frequency of various pathologies such as cardiomegaly, consolidation, edema, lung opacity, nodule/mass, pneumonia, and pneumothorax within the Temporal-MIMIC dataset. The right circular diagram maps the interrelationships among these pathologies where the pathologies are interconnected by colored lines. The thickness of the lines reflect the prevalence of these co-occurrences. \label{fig:class_dist}}
    \end{minipage}
\end{figure*}

We show concept plot on how our dataset is constructed in \Cref{fig:main-dataset-figure}. The demographic distribution of our generated dataset is shown in \Cref{tab:demographics_tab}, where it shows a wide range of race, sex and age and demographics. Additionally, label distribution and their co-occurrences are shown in \Cref{fig:class_dist}.

\section{Additional results}

\subsection{Different multi-modal fusion methods}
\label{sec:fusion_methods}

\begin{itemize}
  \item ViLT \citep{pmlr-v139-kim21k}: We adapt ViLT as general early fusion case for transformer. Given input $\mathbf{x_1}=\mathbb{R}^{L_1\times D}$ and $\mathbf{x_2}\in\mathbb{R}^{L_2\times D}$, where $L_1, L_2$ are sequence length and $D$ is dimension length of representation. $\mathbf{x_1}$ and $\mathbf{x_2}$ are first concatenated with [CLS] token $\mathbf{t}_\text{cls}$ of shape $\mathbb{R}^{1\times D}$ to get input with size $\mathbf{x}=[\mathbf{t}_\text{cls}||\mathbf{x_1}||\mathbf{x_2}]\in\mathbb{R}^{(1+L_1+L_2)\times D}$. Then, the concatenated input is directly feed into a standard transformer model as $\text{Transformer}(\mathbf{x};\theta)$ with learnable tokens $\mathbf{t_1}\in\mathbb{R}^{L_1\times D}$ and $\mathbf{t_2}\in\mathbb{R}^{L_2\times D}$ added to the input along with positional embedding to indicate different modalities. We used 1 layers transformer for early fusion.
  \item MBT \citep{nagrani2021attention}: We adapt MBT as general intermediate fusion case for transformer. Given input $\mathbf{x_1}, \mathbf{x_2}$ following the notation of ViLT. An extra fusion token $\mathbf{x}_\text{fsn}\in\mathbb{R}^{L_3\times D}$ is added in the intermediate layers of transformer such that $\mathbf{x}=[\mathbf{x_1}||\mathbf{x}_\text{fsn}||\mathbf{x_2}]$. Output for each layer of transformer output is then calculated as $[\mathbf{x}_i^{l+1}||\mathbf{\hat{z}}^{l+1}_{\text{fsn}_i}]=\text{Transformer}([\mathbf{x}^{l}_i||\mathbf{z}^l_{\text{fsn}};\theta_{i}])$ given $i$ as index for different modalities. The fusion token is then updated with $\mathbf{z}^{l+1}_\text{fsn}=\text{Avg}_i(\mathbf{\hat{z}}_{\text{fsn}_i}^{l+1})$. We used 6 layers transformer with last 3 layers as intermediate fusion layers in our experiment following the best experiment setting in original paper.
  \item ConcatMLP: For given input $\mathbf{x_1}=\mathbb{R}^{I}$ and $\mathbf{x_2}\in\mathbb{R}^{J}$, where concatenation $\oplus[\mathbf{x_1}, \mathbf{x_2}]\in\mathbb{R}^{I+J}$. Given two weight matrices $\mathbf{W}_1\in\mathbb{R}^{(I+J)\times K}$ and $\mathbf{W}_2\in\mathbb{R}^{K\times M}$ and sigmoid function $\sigma$, fusion output is represented as $\mathbf{y}=\mathbf{W_2}^T\sigma(\mathbf{W_1}^T\oplus[\mathbf{x_1}, \mathbf{x_2}])$
  \item Block \cite{block}: For given bilinear model $\mathbf{y}=\mathbf{\mathcal{T}}\times_1\mathbf{x_1}\times_2\mathbf{x_2}$ with $\mathbf{x_1}\in\mathbb{R}^I, \mathbf{x_2}\in\mathbb{R}^{J}$ and $\mathbf{\mathcal{T}}\in\mathbb{R}^{I\times J\times K}$, Block decomposes the bilinear model on $\mathbf{\mathcal{T}}$ to get form of $\mathbf{y}=\mathbf{C}(\mathbf{\mathcal{D}}\times_1(\mathbf{x_1}^{T}\mathbf{A})\times_2(\mathbf{x_2}^{T}\mathbf{B}))$ with $\mathbf{\mathcal{D}}\in\mathbb{R}^{L\times M\times N}, \mathbf{A}\in\mathbb{R}^{I\times L}, \mathbf{B}\in\mathbb{R}^{J\times M}$ and $\mathbf{C}\in\mathbb{R}^{K\times N}$.
  \item METER \citep{METER}: Two transformer encoders composed of self-attention and cross-attention modules are used with two different modalities as input. The final output $\mathbf{y}$ is computed by a MLP layer on top of concatenation of $[CLS]$ token output of these two transformers. Taking Query, Key, Value of the transformers $Q_1\in\mathbb{R}^{d_k}, K_1\in\mathbb{R}^{d_k}, V_1\in\mathbb{R}^{d_v}$ from one modality and $Q_2\in\mathbb{R}^{d_k}, K_2\in\mathbb{R}^{d_k}, V_2\in\mathbb{R}^{d_v}$ from another modality, the cross-attention mechanism of these two transformers is calculated as $CrossAtt(Q_1, K_2, V_2)=\text{softmax}(\frac{Q_1K_2^T}{\sqrt{d_k}})V_2$ and $CrossAtt(Q_2, K_1, V_1)=\text{softmax}(\frac{Q_2K_1^T}{\sqrt{d_k}})V_1$.
\end{itemize}

\subsection{Per Pathology Model Performance on Different Report Numbers and Time Intervals}
\begin{figure*}[htbp]
\centering
  {\includegraphics[width=1.0\linewidth]{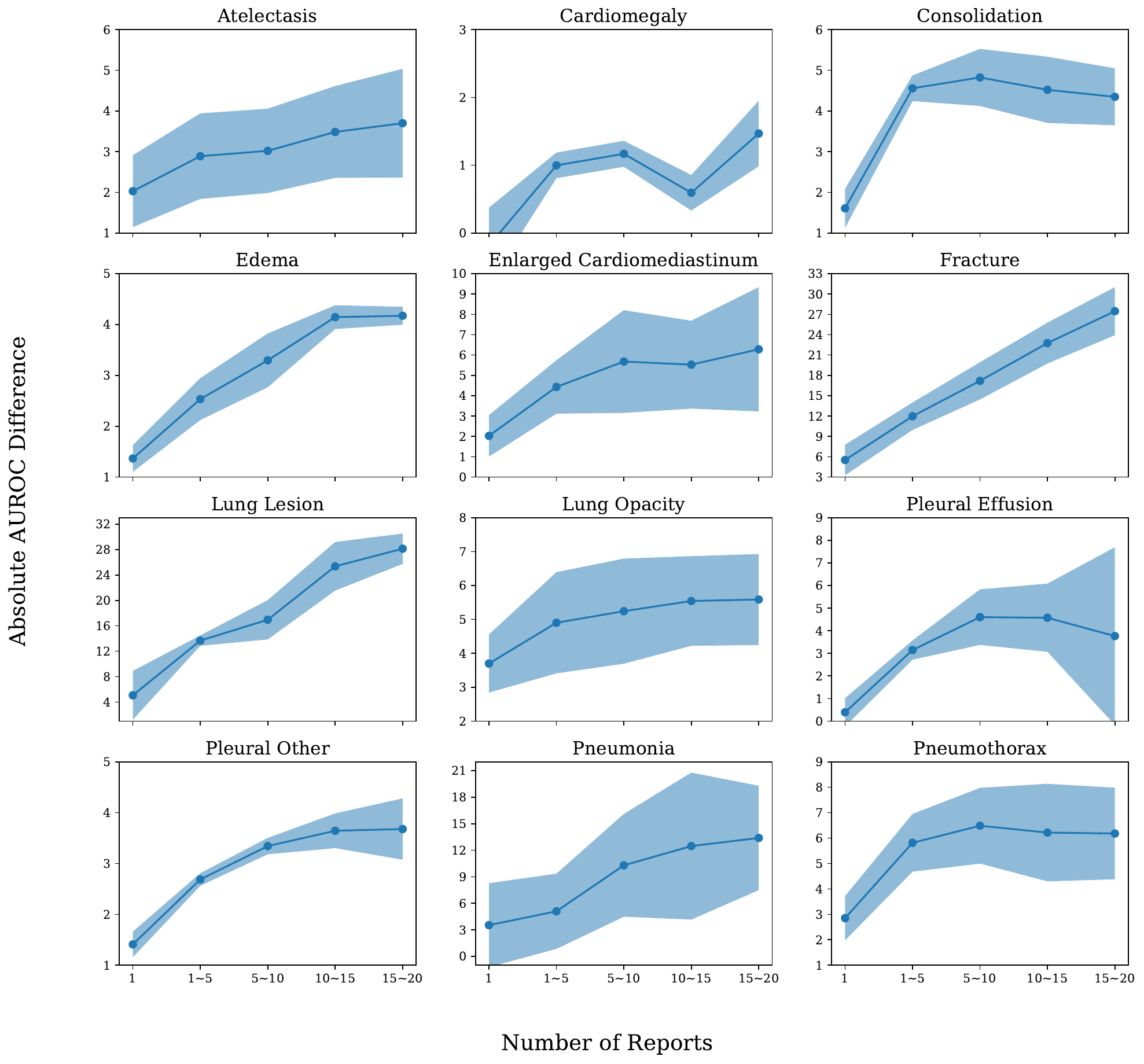}}
  {\caption{Performance Difference of model trained only on current scans in comparison to current scan Images and historical reports on different number of reports for different pathologies.}\label{fig:per_path_num_reports}}
\end{figure*}

\begin{figure*}[htbp]
\centering
  {\includegraphics[width=1.0\linewidth]{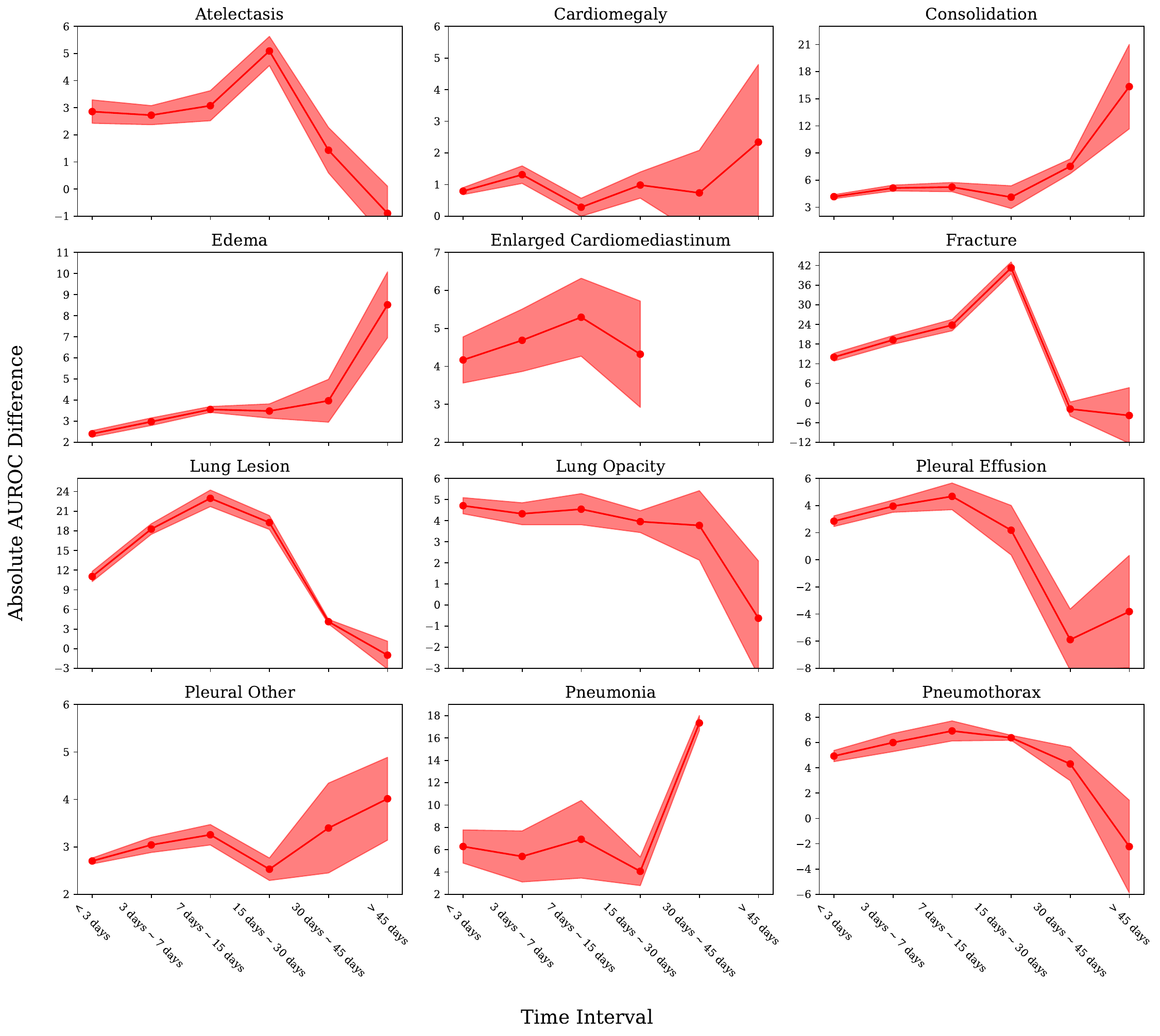}}
  {\caption{Performance Difference of model trained only on current scans in comparison to current scan Images and historical reports on different time intervals for different pathologies.}\label{fig:per_path_time_intervals_reports}}
\end{figure*}

We show all 12 classes performance for number of reports and time intervals ablation in \Cref{fig:per_path_num_reports} and \Cref{fig:per_path_time_intervals_reports} with mean and $95\%$ confidence interval, where we show that the general trend for all pathology in main paper hold for majority of pathologies with few exceptions.

We show additional ablation studies in \Cref{fig:appendix_ablation} to justify our architecture selection. The interpretation for additional three ablation studies are present as follows.

\subsection{Different Position Encoding}
Since previous works have shown that position encoding can have important impact on model performance for transformer-based language model \citep{T5, workshop2022bloom, touvron2023llama, kazemnejad2023the}, we ablate on some popular position encoding in our framework with position to be indicated with time-stamps. The purpose of this ablation is to identify the optimal ways of injecting temporal information to input tokens. In \Cref{fig:appendix_ablation} (a), we compare model performance on different positional encoding, where the definition of sine-cosine, learnable and RoPE \citep{rope} positional encoding are defined in the Supplementary Material. While using no, sine-cosine or learnable positional encoding show similar performance, RoPE shows clear improvement over other methods. This indicates that adding time-series information with relative positional encoding (RoPE) to our architecture can clearly benefit model performance.

\subsection{Different Pooling Methods}
In order to explore the effectiveness of pooling time-series representations with transformer encoder, we compare time-series pooling with transformer (TST) vs. mean pooling (Mean) used in HAIM \citep{Soenksen2022} for evaluating the benefits of pooling with time-series transformer in \Cref{fig:appendix_ablation} (b). The result shows that pooling time-series reports with time-series transformer is more effective than mean pooling when either training with time-series reports along or multi-modal training with both image and reports. When mean pooling is performed, the model is not capable of retaining information interactions across various time-stamps, where time information in each time-stamp is encoded equally. Conversely, the integration of self-attention, complemented by time-stamp positional encoding, enhanced the model's ability to discern and leverage nuanced interactions. This facilitates a more refined and detailed interaction of the representations extracted from different time-stamps. 

We further compare with HAIM~\footnote{\url{https://github.com/lrsoenksen/HAIM}} using feature extraction and fusion with XGBoost \citep{xgboost}. The results indicate suboptimal performance, achieving $\sim61.14$ average AUROC for the image uni-modal, $\sim59.61$ for the text uni-modal, and $\sim63.92$ for the image-text multi-modal model. We attribute this under-performance to the lack of weight updates for the modality-specific models. Further investigation and incorporation of more modalities would be an interesting direction for future work.

\subsection{Different Modality Combinations}
In order to examine the effectiveness of multi-modal fusion in comparison to simplest way of combining different modality with logits averaging, we compare model performance on different modality combination methods in \Cref{fig:appendix_ablation} (c). Ensemble means the logits output from two modality-specific encoders are directly averaged and fusion means early fusion is performed on the representation of two modality-specific encoders. For fair comparison, two encoders trained on different seeds are used for ensembling for image and text result. The result shows that fusion always outperform other ensembling options, showing the effectiveness of our proposed framework.

\subsection{Per Pathology Model Performance on AUROC and AUPRC Plots}
\begin{figure*}[htbp]
\centering
  {\includegraphics[width=0.9\linewidth]{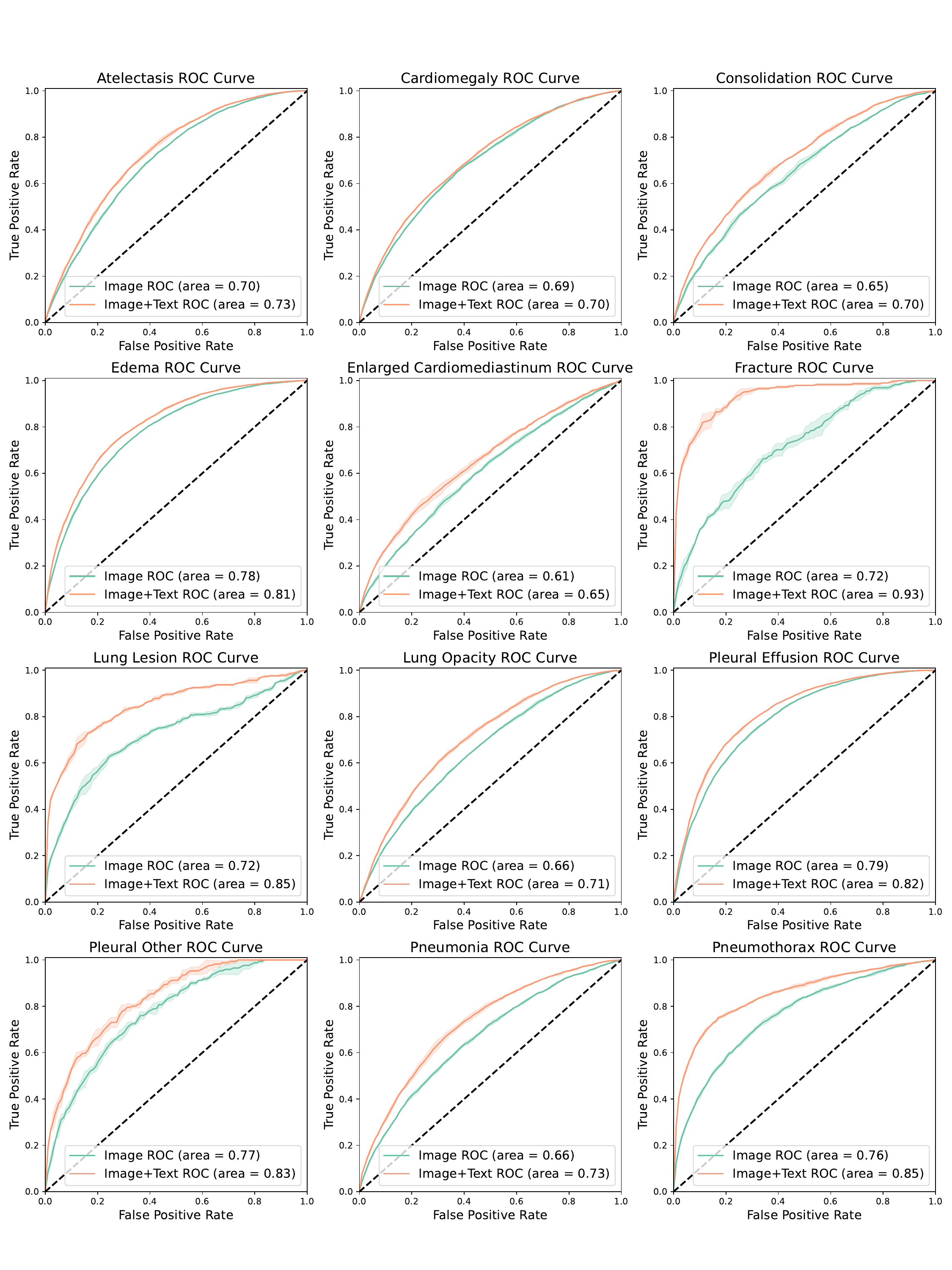}}
  {\caption{Per Pathology ROC Curve}\label{fig:per_path_auroc}}
\end{figure*}

\begin{figure*}[htbp]
\centering
  {\includegraphics[width=0.9\linewidth]{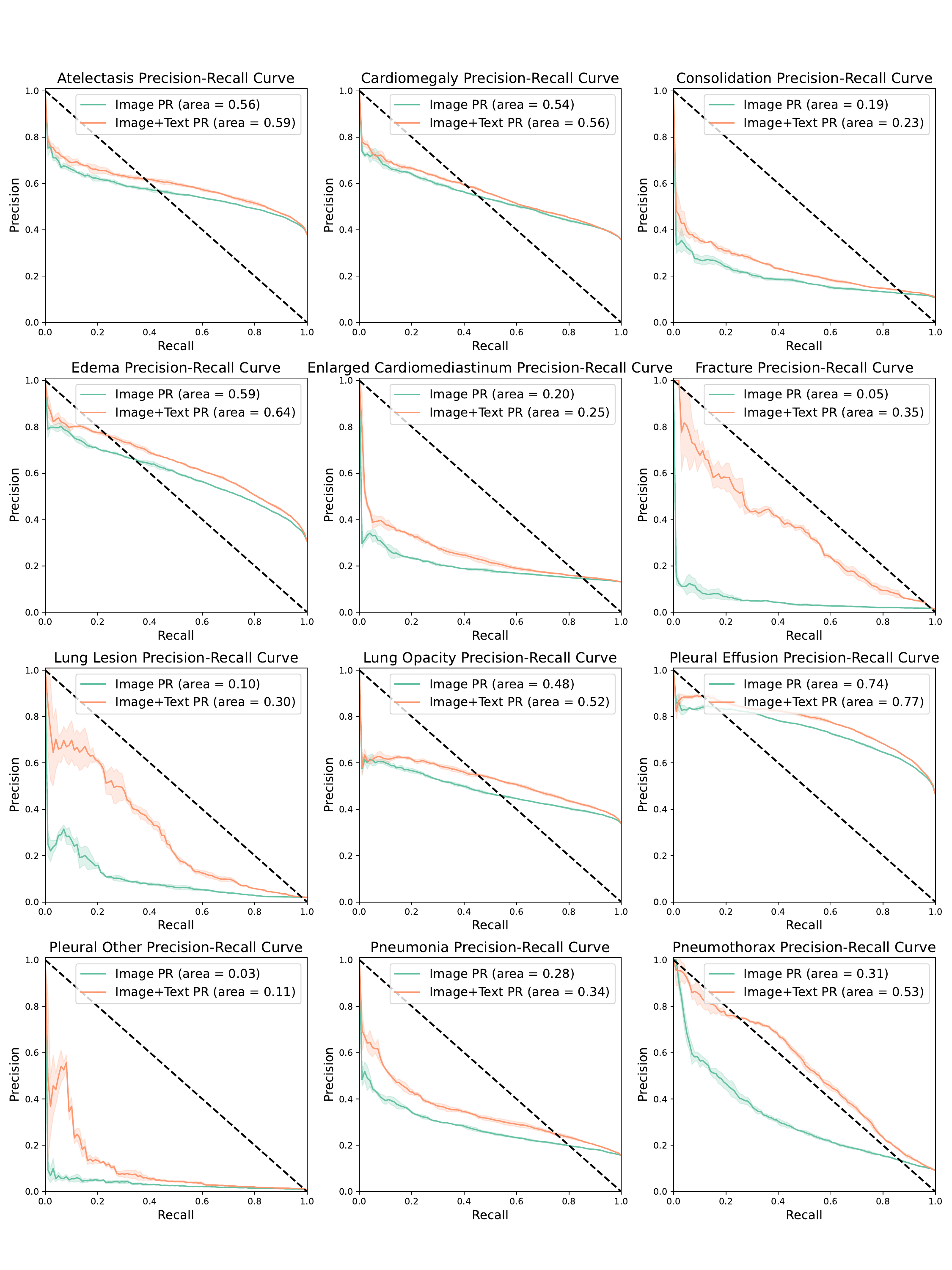}}
  {\caption{Per Pathology Precision-Recall Curve}\label{fig:per_path_auprc}}
\end{figure*}

We show all 12 classes performance with AUROC and AUPRC plots in \Cref{fig:per_path_auroc} and \Cref{fig:per_path_auprc}. In total, there are $4.48\%$ labels multi-modal gets right while image based uni-model gets wrong.

\subsection{Dataset Sample}\label{apd:four}
\begin{figure*}[htbp]
\centering
  {\includegraphics[width=0.8\linewidth]{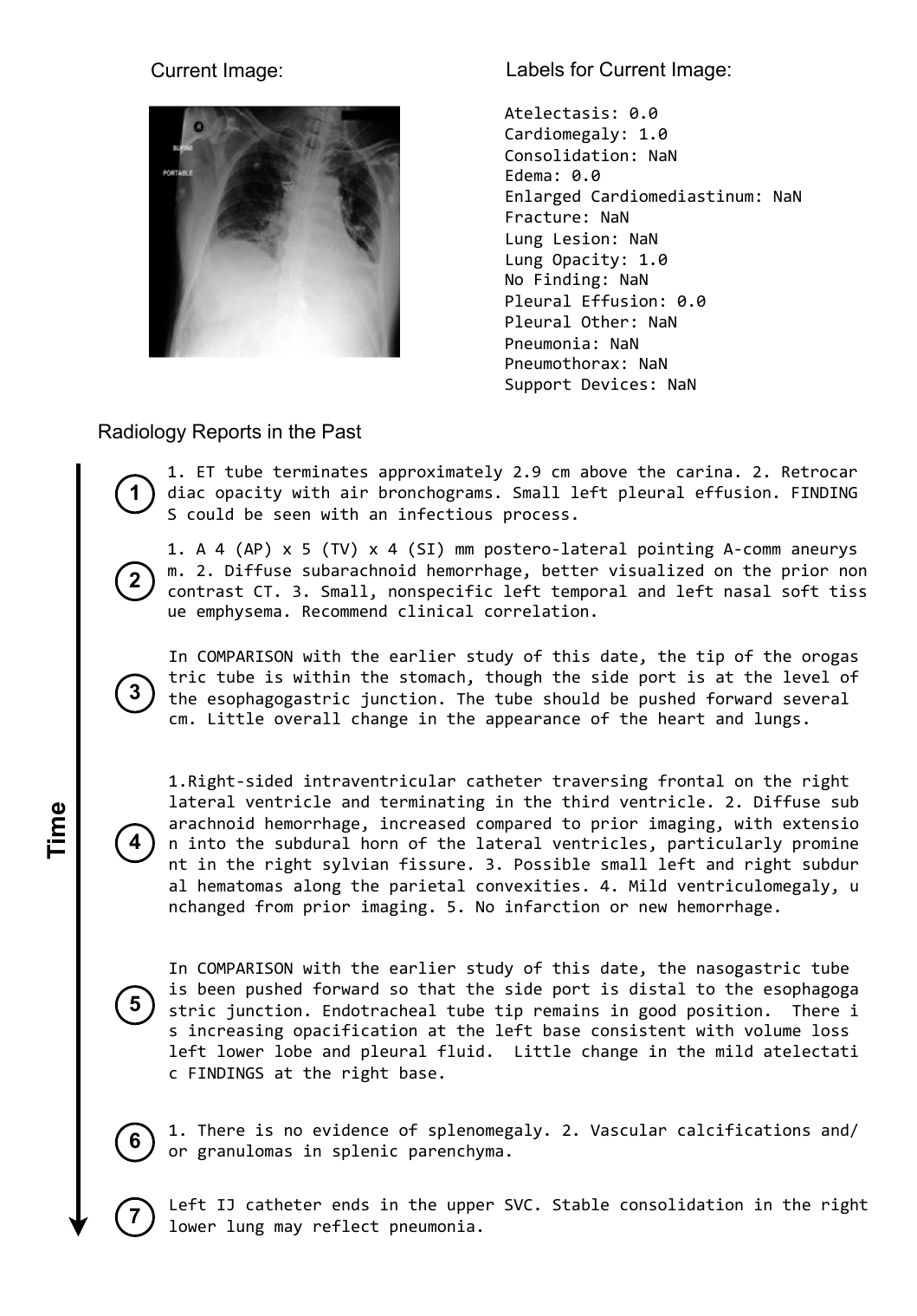}}
  {\caption{\textbf{Dataset Sample:} We show a pre-processed sample from our Temporal MIMIC dataset containing current image and corresponding labels and all previous reports in chronological order.}\label{fig:example}}
\end{figure*}

We show an example of a current radiology image with associated past reports and ground truth in Figure \Cref{fig:example}. 

\input{additional_ablations_figure}

\begin{figure*}[!t]
\centering
  \includegraphics[width=\linewidth]{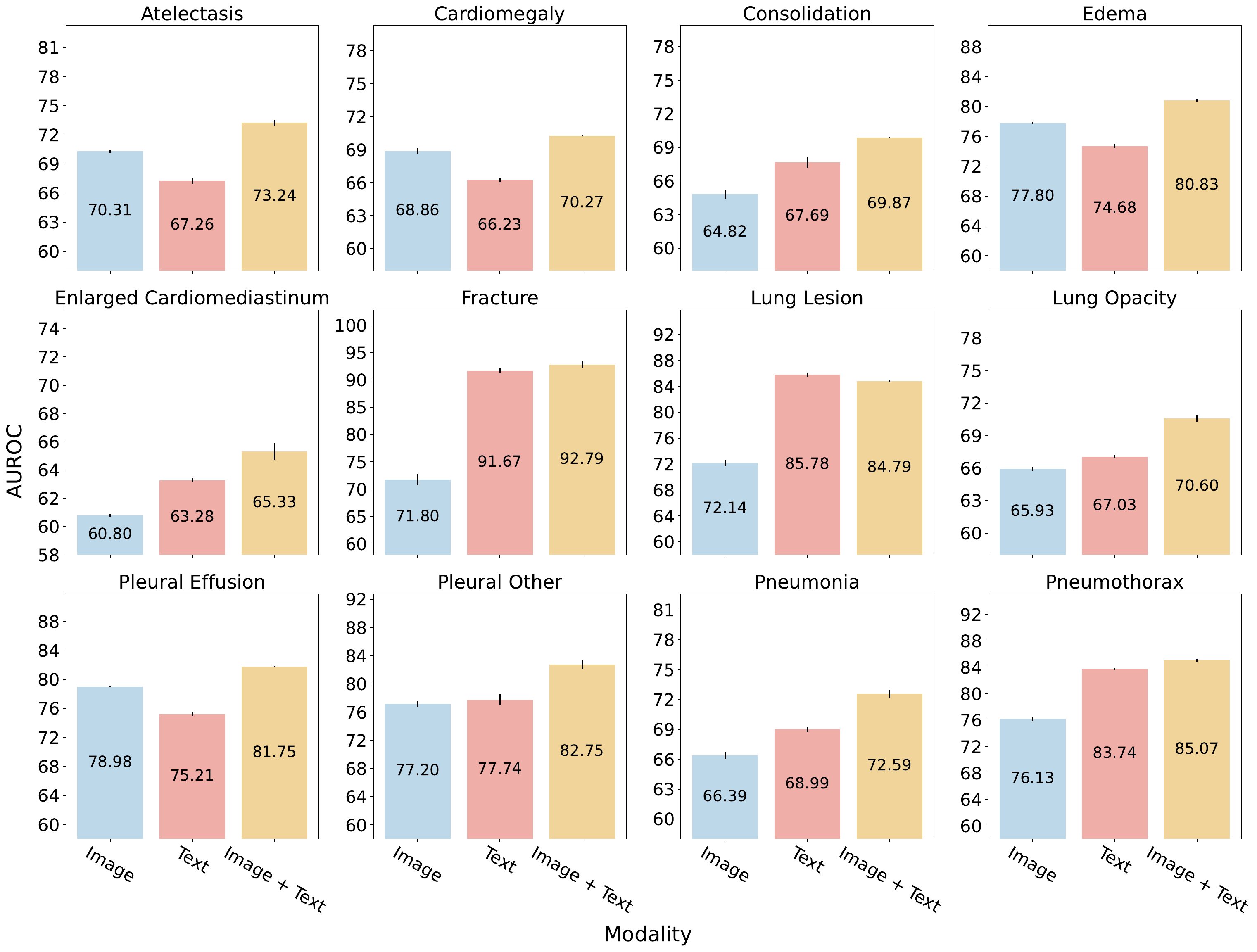}
  \caption{{\bf AUROC comparison between different models for 13 pathologies.} We evaluate the performance of three models (bars in the following order): an {image-only model} in that utilizes images from the current timestamp, {text-only model} that uses reports from previous timestamps, and our comprehensive model that integrates both current images and past textual data for diagnosis. Our model markedly enhances diagnostic accuracy across all examined pathologies.}
  \label{fig:ablation_eval}
\end{figure*}

\end{document}

%% file: results_figure.tex
\begin{figure*}[!ht]
\centering
\begin{tikzpicture}
    \begin{axis}[
        ybar=0pt,
        bar width=.5cm,
        width=\linewidth,
        height=5cm,
        symbolic x coords={Atelectasis, Cardiomegaly, Consolidation, Edema, Cardiomediastinum, Fracture, Lung Lesion, Lung Opacity, Pleural Effusion, Pleural Other, Pneumonia, Pneumothorax},
        xtick=data,
        ymin=55, ymax=95,
        ylabel={AUROC \%},
        nodes near coords,
        nodes near coords align={vertical},
        nodes near coords style={color=black, font=\scriptsize, yshift=0.05cm}, %
        enlarge x limits=0.05,
        legend style={at={(0.5,-0.2)}, anchor=north, legend columns=-1}
        error bars/.cd,
        xticklabel style={rotate=45, anchor=east},
        axis x line*=bottom, 
        axis y line*=left 
        ]
    \addplot[fill=lightblue, draw=none,error bars/.cd, y dir=both, y explicit, 
    error bar style={line width=0.8pt}, 
    error mark options={
      rotate=90,
      black,
      mark size=0pt,
      line width=0pt
    }] coordinates {
        (Atelectasis,70.3) +- (0,0.17) (Cardiomegaly,68.9) +- (0,0.26) (Consolidation,64.8) +- (0,0.38) (Edema,77.8) +- (0,0.14) (Cardiomediastinum,60.8) +- (0,0.09) (Fracture, 71.8) +- (0,1.02) (Lung Lesion,72.1) +- (0,0.47) (Lung Opacity,65.9) +- (0,0.22) (Pleural Effusion,79.0) +- (0,0.10) (Pleural Other, 77.2) +- (0,0.41) (Pneumonia, 66.4) +- (0,0.39) (Pneumothorax, 76.1) +- (0,0.27)};
    \addplot[fill=salmon, draw=none, error bars/.cd, y dir=both, y explicit, 
    error bar style={line width=0.8pt}, 
    error mark options={
      rotate=90,
      black,
      mark size=0pt,
      line width=0pt
    }] coordinates {
        (Atelectasis,73.2) +- (0,0.28) (Cardiomegaly,70.3) +- (0,0.06) (Consolidation,69.9) +- (0,0.06) (Edema, 80.8) +- (0,0.17) (Cardiomediastinum, 65.3) +- (0,0.59) (Fracture,92.8) +- (0,0.62) (Lung Lesion, 84.8) +- (0,0.18) (Lung Opacity, 70.6) +- (0,0.33) (Pleural Effusion, 81.8) +- (0,0.06) (Pleural Other, 82.8) +- (0,0.64) (Pneumonia, 72.6) +- (0,0.38) (Pneumothorax,85.1) +- (0,0.21)};
    \end{axis}
\end{tikzpicture}%

  \caption{{\bf AUROC comparison between different models for 13 pathologies.} We compare the {image-only model} in blue (left) bars that utilizes images from the current timestamp with HIST-AID in red (right), that integrates both current images and past textual data for diagnosis. We show that our model using both current scan and historical reports text enhances AUROC across all pathologies.}
  \label{fig:main_eval}
\end{figure*}

%% file: demographics_figure.tex
\begin{figure*}[!ht]
\begin{tikzpicture}
    \begin{axis}[
        ybar=0pt,
        bar width=.5cm,
        width=0.27\textwidth,
        height=5cm,
        symbolic x coords={Female, Male},
        xtick=data,
        ymin=65, ymax=82,
        ylabel={AUROC \%},
        nodes near coords,
        nodes near coords align={vertical},
        nodes near coords style={color=black, font=\scriptsize, yshift=0.05cm}, %
        enlarge x limits=0.7,
        title={\bf Gender},
        legend style={at={(0.5,-0.2)}, anchor=north, legend columns=-1},
        axis x line*=bottom, 
        axis y line*=left
    ]
    \addplot[fill=lightblue, draw=none, error bars/.cd, y dir=both, y explicit, error bar style={line width=0.8pt}, 
    error mark options={
      rotate=90,
      black,
      mark size=0pt,
      line width=0pt
    }] coordinates {
        (Female,69.5) +- (0,0.10) (Male,71.9) +- (0,0.14)};
    \addplot[fill=salmon, draw=none, error bars/.cd, y dir=both, y explicit, error bar style={line width=0.8pt}, 
    error mark options={
      rotate=90,
      black,
      mark size=0pt,
      line width=0pt
    }] coordinates {
        (Female,76.3) +- (0,0.19) (Male,77.4) +- (0,0.24)};
    \end{axis}
\end{tikzpicture}%
\hfill
\begin{tikzpicture}
    \begin{axis}[
        ybar=0pt,
        bar width=.5cm,
        width=0.4\textwidth,
        height=5cm,
        symbolic x coords={$<$40, 40-60, 60-80, $>$80},
        xtick=data,
        ymin=65, ymax=82,
        ylabel={Mean AUROC \%},
        nodes near coords,
        nodes near coords align={vertical},
        nodes near coords style={color=black, font=\scriptsize, yshift=0.05cm}, %
        enlarge x limits=0.2,
        title={\bf Age},
        legend style={at={(0.5,-0.2)}, anchor=north, legend columns=-1},
        axis x line*=bottom, 
        axis y line*=left
    ]
    \addplot[fill=lightblue, draw=none, error bars/.cd, y dir=both, y explicit, error bar style={line width=0.8pt}, 
    error mark options={
      rotate=90,
      black,
      mark size=0pt,
      line width=0pt
    }] coordinates {
        ($<$40,73.5) +- (0,0.75) (40-60,71.9) +- (0,0.25) (60-80,69.5) +- (0,0.07) ($>$80,68.6) +- (0,0.12)};
    \addplot[fill=salmon, draw=none, error bars/.cd, y dir=both, y explicit, error bar style={line width=0.8pt}, 
    error mark options={
      rotate=90,
      black,
      mark size=0pt,
      line width=0pt
    }] coordinates {
        ($<$40,79.6) +- (0,0.36) (40-60,79.2) +- (0,0.51) (60-80,76.1) +- (0,0.11) ($>$80,72.6) +- (0,0.47)};
    \end{axis}
\end{tikzpicture}%
\hfill
\begin{tikzpicture}
    \begin{axis}[
        ybar=0pt,
        bar width=.5cm,
        width=0.4\textwidth,
        height=5cm,
        symbolic x coords={White, Black, Asian, Other},
        xtick=data,
        ymin=65, ymax=82,
        ylabel={Mean AUROC \%},
        nodes near coords,
        nodes near coords align={vertical},
        nodes near coords style={color=black, font=\scriptsize, yshift=0.05cm}, %
        yshift=-0.6cm,
        enlarge x limits=0.2,
        title={\bf Race},
        legend style={at={(0.5,-0.2)}, anchor=north, legend columns=-1},
        axis x line*=bottom, 
        axis y line*=left
    ]
    \addplot[fill=lightblue, draw=none, error bars/.cd, y dir=both, y explicit, error bar style={line width=0.8pt}, 
    error mark options={
      rotate=90,
      black,
      mark size=0pt,
      line width=0pt
    }] coordinates {
        (White,70.2) +- (0,0.19) (Black,71.2) +- (0,0.51) (Asian,73.8) +- (0,1.31) (Other,72.4) +- (0,0.34)};
    \addplot[fill=salmon, draw=none, error bars/.cd, y dir=both, y explicit, error bar style={line width=0.8pt}, 
    error mark options={
      rotate=90,
      black,
      mark size=0pt,
      line width=0pt
    }] coordinates {
        (White,77.0) +- (0,0.23) (Black,75.6) +- (0,0.19) (Asian,80.9) +- (0,0.63) (Other,78.3) +- (0,0.13)};
    \end{axis}
\end{tikzpicture}
\caption{{\bf AUROC comparison between model trained with current scan and HIST-AID across different demographic groups.} Our model in red (right) consistently outperforms the model in blue (left) trained with current scan images across gender, age groups, and racial categories. The error bars represent standard deviations calculated over five independent runs.}
\label{fig:demo_perf}
\end{figure*}

%% file: ablations_figure.tex
\begin{figure*}[!t]
  \begin{minipage}{0.5\linewidth}
    \centering
  \begin{tikzpicture}
      \centering
    \begin{axis}[
        ybar=0pt,
        bar width=.6cm,
        width=0.85\linewidth,
        height=5cm,
        symbolic x coords={Image, Text},
        xtick=data,
        ymin=60, ymax=75,
        ylabel={Mean AUROC \%},
        nodes near coords,
        nodes near coords style={color=black, font=\scriptsize, yshift=0.05cm}, %
        enlarge x limits=0.7,
        title={\bf Impact of history on uni-modal models},
        legend style={at={(0.5,-0.2)}, anchor=north, legend columns=-1},
        axis x line*=bottom, 
        axis y line*=left
    ]
    \addplot[fill=lightblue, draw=none, error bars/.cd, y dir=both, y explicit, error bar style={line width=0.8pt}, 
    error mark options={
      rotate=0,
      black,
      mark size=0pt,
      line width=0pt
    }] coordinates {
        (Image,70.2) +- (0,0.27) (Text,65.0) +- (0,0.27)};
    \addplot[fill=salmon, draw=none, error bars/.cd, y dir=both, y explicit, error bar style={line width=0.8pt}, 
    error mark options={
      rotate=0,
      black,
      mark size=0pt,
      line width=0pt
    }] coordinates {
        (Image,73.0) +- (0,0.22) (Text,72.9) +- (0,0.19)};
    \end{axis}
  \end{tikzpicture}
  \hfill
   \vspace{-0.01in}
    \resizebox{0.9\linewidth}{!}{%
   
    \begin{tabular}{lc}
        \toprule
        \textbf{Multi-modal models} & \textbf{Mean AUROC \%} \\
        \midrule
        Current scan and last report & $72.6  \pm 0.19$ \\
        Current scan and past reports       & $77.5 \pm 0.17$ \\
        All scans and past reports          & $77.7 \pm 0.02$ \\
        \bottomrule
    \end{tabular}}
    \caption{{\bf Impact of historical images and reports on model performance.} When trained with images and text independently, the model's performance improves significantly with the inclusion of historical images and texts (red bars on the left) compared to using only the current scan and previous report (blue bars on the right). In multi-modal data, historical reports are beneficial, while historical images provide only marginal improvements.\label{fig:hist_abla}}
  \end{minipage}
  \hfill
  \begin{minipage}{0.45\linewidth}
  \begin{tikzpicture}
    \begin{axis}[
        ybar,
        bar width=.7cm,
        bar shift=0pt,
        width=0.8\linewidth,
        height=3.5cm,
        scale only axis,
        clip=false,
        separate axis lines,
        axis on top,
        xmin=0,
        xmax=4,
        xtick={1,2,3},
        x tick style={draw=none},
        xticklabels={Finding, Impression, Both},
        ylabel={Mean AUROC \%},
        enlarge x limits={abs=0.1cm},
        nodes near coords,
        nodes near coords align={vertical},
        nodes near coords style={color=black, font=\scriptsize, 
        yshift=0.05cm}, %
        title={{\bf Impact of section on multi-modal models}},
        legend style={at={(0.5,-0.2)}, anchor=north, legend columns=-1},
        ymin=75, ymax=78.5,
        xticklabel style={rotate=45, anchor=east},
        axis x line*=bottom, 
        axis y line*=left 
      ]
    \addplot[fill=lightblue, draw=none, error bars/.cd, y dir=both, y explicit, error bar style={line width=0.8pt}, 
    error mark options={
      rotate=90,
      black,
      mark size=0pt,
      line width=0pt
    }] coordinates {
        (1,75.9) +- (0,0.17)};
    \addplot[fill=lightyellow, draw=none, error bars/.cd, y dir=both, y explicit, error bar style={line width=0.8pt}, 
    error mark options={
      rotate=90,
      black,
      mark size=0pt,
      line width=0pt
    }] coordinates {
        (2,77.5) +- (0,0.17)};
    \addplot[fill=lightgreen, draw=none, error bars/.cd, y dir=both, y explicit, error bar style={line width=0.8pt}, 
    error mark options={
      rotate=90,
      black,
      mark size=0pt,
      line width=0pt
    }] coordinates {
        (3,78.2) +- (0,0.10)};
    \end{axis}
  \end{tikzpicture}
\caption{{\bf Impact of report sections.} Mean AUROC for our multi-modal framework across all pathologies is reported in this study. We observe that impression is much more effective than finding and using both finding and impression can further improve the model performance when all historical reports are used.\label{fig:sections_abla}}
\end{minipage}
\end{figure*}

%% file: additional_ablations_figure.tex
\begin{figure*}[!t]
    \begin{tikzpicture}
    \begin{axis}[
        ybar,
        bar width=.5cm,
        bar shift=0pt,
        width=0.4\textwidth,
        height=3.5cm,
        scale only axis,
        clip=false,
        separate axis lines,
        axis on top,
        xmin=0,
        xmax=5,
        xtick={1,2,3,4},
        x tick style={draw=none},
        xticklabels={No, Sine-Cosine, Learnable, RoPE},
        ylabel={Mean AUROC \%},
        enlarge x limits={abs=0.05cm},
        nodes near coords,
        nodes near coords align={vertical},
        nodes near coords style={color=black, font=\scriptsize, 
        yshift=0.05cm}, %
        title={\bf (a) Different Positional Encoding},
        legend style={at={(0.5,-0.2)}, anchor=north, legend columns=-1},
        ymin=75, ymax=78.5,
        xticklabel style={rotate=45, anchor=east},
        axis x line*=bottom, 
        axis y line*=left 
      ]
    \addplot[fill=lightblue, draw=none, error bars/.cd, y dir=both, y explicit, error bar style={line width=0.8pt}, 
    error mark options={
      rotate=90,
      black,
      mark size=0pt,
      line width=0pt
    }] coordinates {
        (1,75.7) +- (0,0.15)};
    \addplot[fill=lightyellow, draw=none, error bars/.cd, y dir=both, y explicit, error bar style={line width=0.8pt}, 
    error mark options={
      rotate=90,
      black,
      mark size=0pt,
      line width=0pt
    }] coordinates {
        (2,75.8) +- (0,0.21)};
    \addplot[fill=lightgreen, draw=none, error bars/.cd, y dir=both, y explicit, error bar style={line width=0.8pt}, 
    error mark options={
      rotate=90,
      black,
      mark size=0pt,
      line width=0pt
    }] coordinates {
        (3,75.9) +- (0,0.27)};
    \addplot[fill=salmon, draw=none, error bars/.cd, y dir=both, y explicit, error bar style={line width=0.8pt}, 
    error mark options={
      rotate=90,
      black,
      mark size=0pt,
      line width=0pt
    }] coordinates {
        (4,77.5) +- (0,0.17)};
    \end{axis}
    \end{tikzpicture}
  \hfill
    \begin{tikzpicture}
    \begin{axis}[
        ybar=0pt,
        bar width=.5cm,
        width=0.4\textwidth,
        height=5cm,
        symbolic x coords={Image, Text, Multi-Modal},
        xtick=data,
        xticklabel style={rotate=45, anchor=east},
        ymin=65, ymax=80,
        ylabel={Mean AUROC \%},
        nodes near coords,
        nodes near coords align={vertical},
        nodes near coords style={color=black, font=\scriptsize, yshift=0.05cm}, %
        enlarge x limits=0.3,
        title={\bf (b) Different Pooling Methods},
        legend style={at={(0.5,-0.2)}, anchor=north, legend columns=-1},
        axis x line*=bottom, 
        axis y line*=left
    ]
    \addplot[fill=lightblue, draw=none, error bars/.cd, y dir=both, y explicit, error bar style={line width=0.8pt}, 
    error mark options={
      rotate=90,
      black,
      mark size=0pt,
      line width=0pt
    }] coordinates {
        (Image,70.5) +- (0,0.42) (Text,69.3) +- (0,0.57) (Multi-Modal,76.3) +- (0,0.19)};
    \addplot[fill=salmon, draw=none, error bars/.cd, y dir=both, y explicit, error bar style={line width=0.8pt}, 
    error mark options={
      rotate=90,
      black,
      mark size=0pt,
      line width=0pt
    }] coordinates {
        (Image,72.3) +- (0,0.28) (Text,72.5) +- (0,0.21) (Multi-Modal,77.5) +- (0,0.17)};
    \end{axis}
    \end{tikzpicture}
    \hfill 
    \begin{tikzpicture}
    \begin{axis}[
        ybar,
        bar width=.5cm,
        bar shift=0pt,
        width=0.4\textwidth,
        height=3.5cm,
        scale only axis,
        clip=false,
        separate axis lines,
        axis on top,
        xmin=0,
        xmax=5,
        xtick={1,2,3,4},
        x tick style={draw=none},
        xticklabels={Image-Avg, Text-Avg, Both-Avg, Both-Fuse},
        ylabel={Mean AUROC \%},
        enlarge x limits={abs=0.05cm},
        nodes near coords,
        nodes near coords align={vertical},
        nodes near coords style={color=black, font=\scriptsize, 
        yshift=0.05cm}, %
        title={\bf (c) Different Modality Combination},
        legend style={at={(0.5,-0.2)}, anchor=north, legend columns=-1},
        ymin=68, ymax=78.5,
        xticklabel style={rotate=45, anchor=east},
        axis x line*=bottom, 
        axis y line*=left 
      ]
    \addplot[fill=lightblue, draw=none, error bars/.cd, y dir=both, y explicit, error bar style={line width=0.8pt}, 
    error mark options={
      rotate=90,
      black,
      mark size=0pt,
      line width=0pt
    }] coordinates {
        (1,71.2) +- (0,0.11)};
    \addplot[fill=lightyellow, draw=none, error bars/.cd, y dir=both, y explicit, error bar style={line width=0.8pt}, 
    error mark options={
      rotate=90,
      black,
      mark size=0pt,
      line width=0pt
    }] coordinates {
        (2,74.3) +- (0,0.12)};
    \addplot[fill=lightgreen, draw=none, error bars/.cd, y dir=both, y explicit, error bar style={line width=0.8pt}, 
    error mark options={
      rotate=90,
      black,
      mark size=0pt,
      line width=0pt
    }] coordinates {
        (3,76.3) +- (0,0.15)};
    \addplot[fill=salmon, draw=none, error bars/.cd, y dir=both, y explicit, error bar style={line width=0.8pt}, 
    error mark options={
      rotate=90,
      black,
      mark size=0pt,
      line width=0pt
    }] coordinates {
        (4,77.5) +- (0,0.17)};
    \end{axis}
    \end{tikzpicture}
\vspace{-0.1in}
\caption{{\bf Additional Ablation Studies.} {\bf a) For positional encoding ablation}, we evaluate four most commonly used positional encoding, where we find using relative encoding with RoPE consistently perform better. {\bf b) For pooling method ablation}, we compared mean pooling (blue bars) and time-series transformer pooling (red bars), where time-series transformer pooling consistent outperform mean pooling. {\bf c) For different modality combination comparison}, we compared ensembling with logits averaging (blue, yellow and green bars) vs. early fusion (pink bar), where early fusion consistently performs better than logits averaging from two models of experimented modalities.}
\label{fig:appendix_ablation}
\end{figure*}
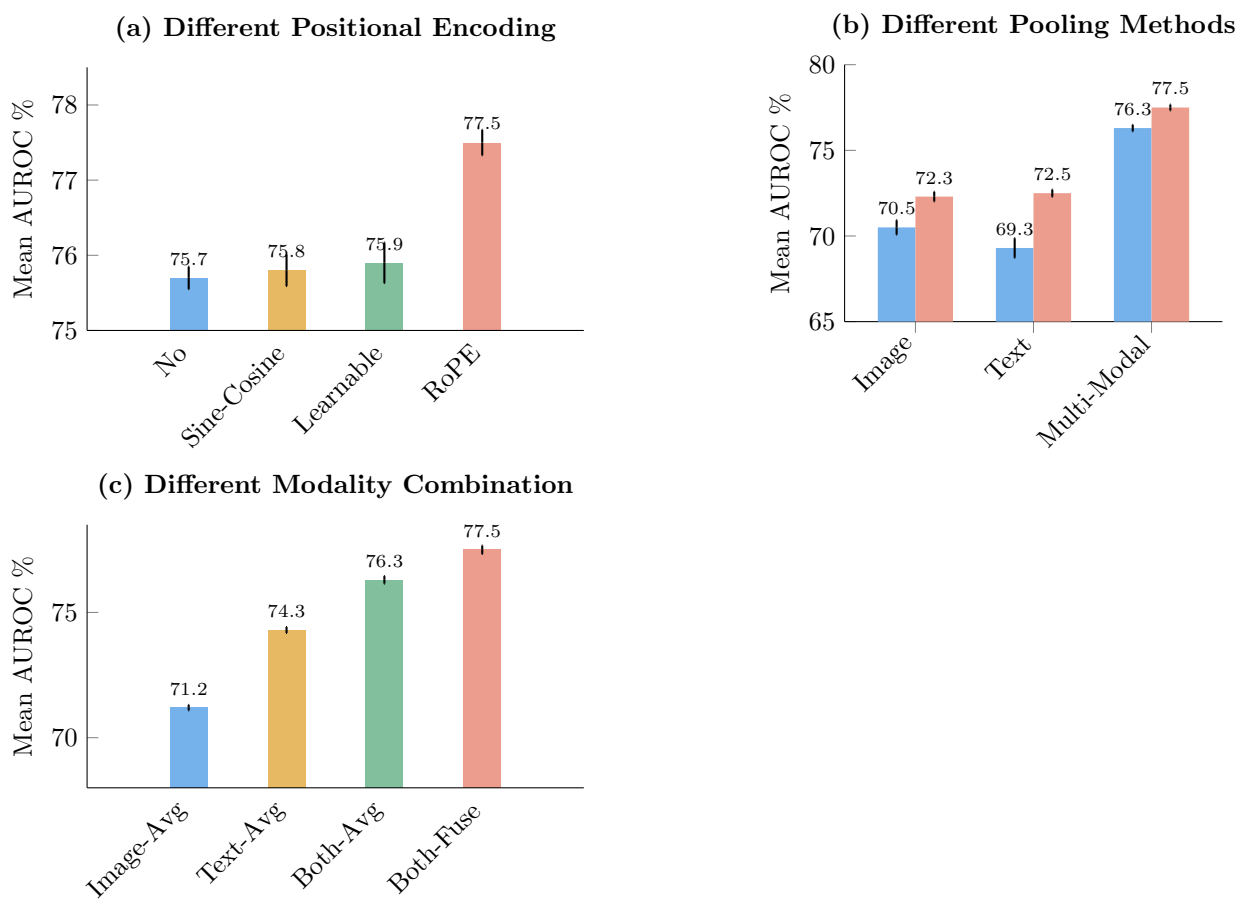